\documentclass[aps,pre,reprint,amsmath,amssymb, showkeys, superscriptaddress]{revtex4-1} 
\usepackage{color,soul} 
\usepackage{mathtools}
\usepackage{float}
\usepackage{graphicx}
\usepackage{amsmath}
\usepackage{amsthm}
\usepackage{amssymb}
\usepackage{multirow} 
\usepackage[normalem]{ulem}
\usepackage[justification=raggedright]{caption}
\usepackage{subcaption}
\usepackage{wrapfig}
\graphicspath{{./images/}}

\begin{document}

\title{Derivation of a statistical model for classical systems obeying the fractional exclusion principle}

\author{Projesh Kumar Roy}
\email{projeshkr@imsc.res.in}
\affiliation{The Institute of Mathematical Sciences, 4th cross road, C.I.T. Campus, Taramani, Chennai -- 600113, Tamil Nadu, India}
\affiliation{Homi Bhabha National Institute, Training School Complex, Anushakti Nagar, Mumbai -- 400094, Maharashtra, India}
\date{\today}
            
\begin{abstract}
The violation of the Pauli principle has been surmised in several models of the Fractional Exclusion Statistics and successfully applied to several quantum systems. In this paper, a classical alternative of the exclusion statistics is studied using the maximum entropy methods. The difference between the Bose-Einstein statistics and the Maxwell-Boltzmann statistics is understood in terms of a separable quantity, namely the degree of indistinguishability. Starting from the usual Maxwell-Boltzmann microstate counting formula, a special restriction related to the degree of indistinguishability is incorporated using Lagrange multipliers to derive the probability distribution function at equilibrium under NVE conditions. It is found that the resulting probability distribution function generates real positive values within the permissible range of parameters. For a dilute system, the probability distribution function is intermediate between the Fermi-Dirac and Bose-Einstein statistics and follows the exclusion principle. Properties of various variables of this novel statistical model are studied and possible application to classical thermodynamics is discussed. 
\end{abstract}


\maketitle

\section{Introduction}

It is known for a long time that, the Pauli exclusion principle is essentially a limiting case in the quantum theory of fundamental particles and not always valid in certain scenarios~\cite{Milotti_Zmeskal_IntJModPhysA_2007}. It was understood, several systems in nature, e.g. 2D-anyone, follow a {\it mixed} statistics which can not be explained using any existing fundamental statistical theories. In the past century, various attempts were made to unify the three fundamental statistical theories present in nature -- namely Maxwell-Boltzmann statistics (MB), Fermi-Dirac statistics (FD), and Bose-Einstein statistics (BE) -- under a single statistical theory which will be able to explain the mixed statistics phenomenon. Gentile statistics~\cite{Gentile_NuovoCimento_1940}, Tsallis statistics~\cite{Tsallis_JStatPhys_1988}, collision model~\cite{March_Sung_PhysChemLiq_1993, March_PhysChemLiq_1997}, Acharya-Swamy statistics~\cite{Acharya_Swamy_JPhysA_1994} etc. were some of the early attempts to unify all fundamental statistical theories. In this context, the theory of {\it fractional exclusion statistics} (FES)~\cite{Khare_2005} was developed based on {\it fractional exclusion principle} (FEP), which was designed to be intermediate between FD and BE statistics. FEP refers to the multiple-but-finite occupancy of the building blocks at a certain state. In this paper, the term \textit{particle} is used to describe the {\it building blocks} of any realistic natural system. It has important applications in the many quantum systems like quantum computation~\cite{Nayak_Sarma_RevModPhys_2008}, anyone statistics~\cite{Veigy_Ovury_PhysRevLett_1994, Veigy_Ovury_ModPhysLettA_1995_1, Veigy_Ovury_ModPhysLettA_1995_2, Bartolomei_Feve_Science_2020}, fractional quantum hall effect~\cite{Stern_AnnPhys_2008}, high temperature superconductivity~\cite{Laughlin_PhysRevLett_1988, Laughlin_Science_1988}, even black holes~\cite{Abchouyeh_Younesizadeh_PhysLettB_2018}. Most of the early versions of FES models (described above) were not derived from first-principle-based reasoning, but mostly based on empirical extrapolations between FD and BE statistics. In 1994, Wu~\cite{Haldane_PhysRevLett_1991, Wu_PhysRevLett_1994, Nayak_Wilczek_PhysRevLett_1994, Anghel_JPhysA_2007, Anghel_EurPhysLett_2009, Anghel_JPhysConfSer_2013} and Isakov~\cite{Isakov_PhysRevLett_1994, Isakov_ModPhysLettB_1994} independently proposed a novel FES model -- namely Haldane-Wu statistics (HW) -- which was derived directly from counting the number of microstates for a strongly correlated quantum system via maximum entropy methods~\cite{Jaynes_PhysRev_1957, Jaynes_PhysRev_1957_2} (MaxEnt). HW statistics aims to compute the correct number of conformations of a system evolving under FEP. It provides a microscopic view of the particle distribution at various energy levels and generalizes the Pauli exclusion principle to multiple occupation numbers. It should be noted, a similar (but not identical) model to HW statistics was proposed by Ramanathan~\cite{Ramanathan_PhysRevD_1992} in 1992, also based on modified microstates counting.

HW statistics has been successful in predicting the statistics of 2D-anyone systems projected at lowest Landau level under a strong magnetic field~\cite{Veigy_Ovury_ModPhysLettA_1995_1, Veigy_Ovury_ModPhysLettA_1995_2} and excitation in pure Laughlin liquids~\cite{Camino_Goldman_PhysRevB_2005, Arovas_Wilczek_PhysRevLett_1984}. Not just quantum systems, HW statistics have significant applications in realistic classical systems as well, e.g., adsorption of polyatomic molecules onto substrates~\cite{Riccardo_Roma_PhysRevLett_2004, Riccardo_Pastor_AppSurfSci_2005, Cerofolini_JPhysA_2006, Davila_Pastor_SurfSci_2009, Davila_Pastor_JChemPhys_2009, Fernandez_Pastor_Langmuir_2011, Fernandez_Pastor_ChemPhysLett_2014, Riccardo_Pasinetti_PhysRevLett_2019}, yielding phenomenon under mechanical stress~\cite{Mezzasalma_RSCAdv_2019}, calculation of entropy in ice~\cite{Ihm_JPhysA_1996}, coil-helix transition transition of polypeptide~\cite{Sharma_Muller_JStatMech_2015}, etc. However occurrence of negative probabilities is a problem in HW statistics~\cite{Nayak_Wilczek_PhysRevLett_1994, Polychronakos_PhysLettB_1996, Polychronakos_NuclPhysB_1996, Chaturvedi_Srinivasan_PhysicaA_1997, Chaturvedi_Srinivasan_PhysRevLett_1997, Murthy_Shankar_PhysRevB_1999}. As pointed out by A. Polychronakos~\cite{Polychronakos_PhysLettB_1996}, this problem is inherent in HW statistics and cannot be avoided via any approximation. An alternate statistical model was suggested by him using path integral formulation -- namely Polychronakos statistics (AP)~\cite{Polychronakos_PhysLettB_1996, Polychronakos_NuclPhysB_1996, Chung_Hassanabadi_ModPhysLettB_2018, Hoyuelos_PhysicaA_2018} -- which avoids the problem of negative probabilities under certain conditions. Following the success of HW and AP statistics, several other FES models based on microstate counting have also been suggested, e.g. P\'olya-urn model~\cite{Niven_EurPhysJB_2009, Niven_Grendar_PhysLettA_2009}, Fractional Superstatistics~\cite{Ourabah_Tribeche_PhysRevE_2018} model, and other mixed statistics models~\cite{Abutaleb_IntJTheoPhys_2014, Yan_PhysRevE_2021}.

Parallel to Wu, Kaniadakis et al.~\cite{Kaniadakis_Quarati_PhysRevE_1993, Kaniadakis_Quarati_PhysRevE_1994, Kaniadakis_PhysLettA_1995, Kaniadakis_Quarati_NuclPhysB_1996} developed a classical analog of FES -- namely the Kinetic model -- by including the FEP in the non-linear Fokker-Plank equation. In this kinetic model, only classical particles are considered; however, their hopping into different states was constrained using the FEP. In this way, quantum effects were introduced into the dynamics of a classical system. This model ultimately evolved into {\it $\kappa$-deformed} statistics~\cite{Kaniadakis_PhysicaA_2001, Kaniadakis_Scarfone_PhysicaA_2002, Kaniadakis_Scarfone_Entropy_2019}, which was able to reproduce both HW and AP type statistics depending on the particular choice of the transition parameters~\cite{Kaniadakis_Quarati_NuclPhysB_1996, Kaniadakis_PhysicaA_2001}. 

All FES models described above have originated from quantum statistics. Although quantum-FES (QFES) models have been widely used in certain classical systems (described above), such strict validity of FEP in classical systems can not be ascertained {\it a priori}. By definition, the microstate counting formula of the classical systems cannot be different from the MB statistics, but the total entropy of the system can be amended using additional constraints. Using this idea, an alternate FES-type model for classical systems is provided -- namely the classical fractional exclusion statistics (CFES) -- without {\it a priori} assumption of the validity of FEP. The idea of generating quantum effects in classical systems is investigated by understanding the difference between the two limiting cases -- {\it i.} completely distinguishable MB statistics, {\it ii.} completely indistinguishable BE statistics. It is shown that the FEP -- much similar to the QFES models -- emerges naturally under certain constraints in classical systems obeying the CFES model. 

The paper is summarized as follows -- section~\ref{sect:DI}: the degree of indistinguishability is formulated from BE microstate equation, section~\ref{sect:cfes}: the principal equations of CFES is derived, section~\ref{sect:properties}: the origin of FEP and other properties of CFES is described, section~\ref{sect:thermodynamics}: application of CFES in classical thermodynamics is discussed.

\section{Degree of Indistinguishability}
\label{sect:DI}

It was understood that quantum effects can be generated by controlling the {\it degree of indistinguishability} (DI) of the particles in the system~\cite{Kaniadakis_Quarati_PhysRevE_1994}. The physical meaning of varying indistinguishability can be that some particles of the same {\it species} (i.e. similar physical traits) can be made distinguishable using some unknown traits which cannot be incorporated in the model. Based on the idea of varying indistinguishability, Medvedev~\cite{Medvedev_PhysRevLett_1997, Meljanac_Ristic_ModPhysLettA_1999} developed another QFES model -- namely {\it ambiguous} statistics -- which is intermediate between FD and BE statistics.

Let's consider a system of total $N$ \textit{non-interacting} particles have a population of $n_i \geq 0$ at $i^{\text{th}}$ energy level $\epsilon_i \geq 0$ with degeneracy $g_i \geq 1$ under NVE condition with total energy $E=\sum_in_i\epsilon_i$ and total degeneracy $G=\sum_ig_i$. The reduced unit formalism is followed in all subsequent sections, i.e., $k_B=1$, where $k_B$ is the Boltzmann constant. For indistinguishable particles, the number of microstates can be calculated from the BE statistics. Using Stirling's approximation $n_i!\sim n_i^{n_i}\exp(-n_i)$ and $n_i << g_i$, one can simplify BE equation for total number microstates, $W_{\text{BE}}$, as,

\begin{equation}
    W_{\text{BE}} \approx W_{\text{MB}} \Bigg [ \frac{1}{N!} \prod_i \Big (1+\frac{n_i}{g_i} \Big )^{n_i} \Bigg ]
    \label{eqn:microstates}
\end{equation}

For details of the derivation, see supporting information section SI.I. The first product of the final Equation~\ref{eqn:microstates} is the traditional microstate counting from MB statistics; $W_{\text{MB}}$ = $\prod_i N!(g_i^{n_i}/n_i!)$; for distinguishable particles. Equation~\ref{eqn:microstates} is usually reduced to the traditional MB statistics by considering $n_i/g_i \sim 0$ -- i.e. {\it dilute} systems -- and the second product $(1+{n_i}/{g_i} )^{n_i} ~ \sim 1$. However, by not neglecting the second product, we get an interesting interpretation of the BE statistics. If the entropy of a BE system is defined as $S_{\text{BE}}=\ln(W_{\text{BE}})$, then using Stirling's approximation and logarithmic expansion on Equation~\ref{eqn:microstates} (see supporting information section SI.I), one can write,

\begin{equation}
    S_{\text{BE}} = S_{\text{MB}} + \Bigg [ \sum_{m=0}^{\infty} f_mS_m - (N\ln N + G)\Bigg ]
    \label{eqn:BE_entropy}
\end{equation}

where $f_m$'s are the coefficients from the logarithmic expansion. Equation~\ref{eqn:BE_entropy} can be viewed as a relationship between the entropic contributions arising from the distinguishable ($S_{\text{MB}}$) and indistinguishable ($S_{\text{BE}}$) properties of the system with unlimited occupancy at each energy level. The residual entropic term, $S^*=\sum_{m=0}^{\infty} f_mS_m - (N\ln N + G)$, can be viewed as a superposition of multiple entropic terms, $S_m$'s, where $ 0 \leq m \leq \infty$. The $\sum_m f_m S_m$ term controls the transition between the distinguishable and indistinguishable particles, hence, one can identify this term as the DI. Maximizing $S_{\text{BE}}$ with respect to the population at each energy state, $n_i$, Equation~\ref{eqn:BE_entropy} becomes,

\begin{equation}
    \Bigg(\frac{\partial S_{\text{BE}}}{\partial n_i} \Bigg)_{\text{max}} = \Bigg(\frac{\partial S_{\text{MB}}}{\partial n_i}\Bigg)_{\text{max}} + \sum_{m=0}^{\infty} f_m \Bigg ( \frac{\partial S_m}{\partial n_i} \Bigg )_{\text{max}} = 0
    \label{eqn:BE_maxen}
\end{equation}

Both $S_{\text{MB}}$ and $S_m$ terms and their derivatives are real positive quantities. Hence, each individual $S_m$ have to be constant at the maximum value of $S_{\text{BE}}$. This result can be stated in a different way as, ($S_m$ = {\it constant}) constraint acts as an additional restriction on top of ($S_{\text{MB}}$ = {\it constant}) to generate the maximum entropy in BE statistics for indistinguishable particles. However, determination of the effects of individual $S_m$ values are not required to get $S_{\text{BE}}$ from $S_{\text{MB}}$, as their linear superposition with coefficients $f_m$ leads to an $m$ independent logarithmic function in Equation~\ref{eqn:BE_entropy}. Question is, how will the statistics of the system evolve if the superposition of each of these restrictions does not lead to an $m$ independent function.

To answer such a question, one has to understand how a particle statistics evolves under each of these restrictions -- i.e. ($S_m$ = {\it constant}) -- in addition to ($S_{\text{MB}}$ = {\it constant}). A statistical model can be prepared which maximize $S_m$ in addition to $S_{\text{MB}}$. MaxEnt method~\cite{Jaynes_PhysRev_1957, Jaynes_PhysRev_1957_2} is suitable for introducing any additional restriction in a statistical model using Lagrange multipliers. One can expect that, properties of such a model should lie in between BE and MB statistics. 

\section{A Classical Fractional Exclusion Statistics}
\label{sect:cfes}

Taking inspiration from Equation~\ref{eqn:BE_entropy}, the following relation between $n_i$ and $g_i$ is used to control DI in a system of distinguishable particles as,

\begin{equation}
     \sum_i \frac{n_i^m}{g_i^{(m-1)}} = S_m
    \label{eqn:conditions_1}        
\end{equation}

where, $m$ is defined as $m \in \mathbb{R}$, and $S_m$ is a constant. The aim of the following analysis is to study the effect of different $m$ values on $S_{\text{MB}}$ using MaxEnt method. For $m$=0 or 1, Equation~\ref{eqn:conditions_1} generates the trivial conditions for NVE ensemble. For Equation~\ref{eqn:conditions_1} to be an non-trivial constraint, $S_m$ cannot be linearly dependent on $N$, $E$, or $G$, and must be an independent external parameter. Later it is shown that, $S_m$ is related to an effective particle number. Therefore, starting from a purely MB type statistics, CFES incorporates the individual effects of the constraints ($S_m$={\it constant}) from Equation~\ref{eqn:BE_entropy} on the maximum value of $S_{\text{MB}}$. It can be shown that, one cannot use any other exponent combinations on $n_i$ and $g_i$ in Equation~\ref{eqn:conditions_1} (see supporting information section SI.II). 

The derivation of CFES considers the absolute number of microstates to be same as used in the modified MB statistics as, $\widetilde{W} = W_{\text{MB}}/N!$. At equilibrium, one can write the optimal distribution of the particles corresponding to the maximum entropy as (see supporting information section SI.III),

\begin{equation}
    n_i = g_i e^{-(\alpha + \beta \epsilon_i)} e^{-m\gamma(n_i/g_i)^{(m-1)}}
    \label{eqn:CFES_1}
\end{equation}

where $\alpha$, $\beta$, and $\gamma$ are the Lagrange multipliers. If one uses the approximation that in dilute systems $g_i >> n_i$, then

\begin{equation}
    \frac{n_i}{g_i} = e^{-(\alpha+\beta\epsilon_i)}\Bigg [ 1 - m\gamma \Bigg (\frac{n_i}{g_i} \Bigg )^{(m-1)} \Bigg ]
    \label{eqn:CFES_2}
\end{equation}

Equation~\ref{eqn:CFES_2} is the final expression of the probability distribution function in the CFES. $\exp(-\alpha) \equiv N/Z_m$ is the normalization factor and $Z_m$ is the partition function corresponding to Equation~\ref{eqn:CFES_2}. The average population per individual degenerate levels with energy $\epsilon_i$ is defined as $\bar{n_i}=(n_i/g_i)$. Equation~\ref{eqn:CFES_2} can be reduced using Ramanujan formula~\cite{Berndt_Wilson_AdvMath_1983, Murthy_Shankar_PhysRevB_1999}. Details of the derivation is given in supporting information section SI.IV. The reduced solution of Equation~\ref{eqn:CFES_2} for $m \geq 2$ can be written as,

\begin{equation}
    \bar{n_i} = \sum_{j=0}^{\infty} C_j(m) (-m\gamma)^j e^{-[(mj-j+1)(\alpha+\beta\epsilon_i)]}
    \label{eqn:CFES_3}
\end{equation}

where $C_j(m)$ is the first weight to the probability function. It is defined as,

\begin{equation}
C_j(m) = \frac{1}{j!}\prod_{k=1}^{j-1} [1 + j(m-1) -k] = \prod_{l=2}^j \Bigg( \frac{(m-2)j}{l} + 1 \Bigg)
\label{eqn:CFES_weight}
\end{equation}

From the definition of $C_j(m)$, it is clear that it is always a positive number for $m \geq 2$. The second weight, $(-m\gamma)^j$, is always positive for $\gamma < 0$ and $m > 0$. For $\gamma > 0$ and $m > 0$, this factor can either be positive or negative, depending on whether $j$ is odd or even. If the total weights are negative at a certain $(m,\gamma)$ combination, the total probability function will become negative as well; which is a common problem in HW statistics. Later it is shown that, within the permissible range of $m$ and $\gamma$, negative probabilities do not arise in CFES.  
    
For $m=0$, it is straight forward to show that Equation~\ref{eqn:CFES_2} becomes traditional MB distribution. If $m=1$, then Equation~\ref{eqn:CFES_2} becomes,

\begin{equation}
    \bar{n_i} = e^{-(\alpha+\beta\epsilon_i)}(1-\gamma)
    \label{eqn:Order1}
\end{equation}

Equation~\ref{eqn:Order1} is also equivalent to the traditional MB distribution. Via normalizing, one can find $\exp(-\alpha) = N/[Z_{\text{MB}}(1-\gamma)]$, where $Z_{\text{MB}} = \sum_i g_i\exp(-\beta \epsilon_i)$ is the partition function corresponding to the  MB statistics. If $m=2$, then Equation~\ref{eqn:CFES_2} becomes,

\begin{equation}
    \bar{n_i} = \frac{1}{e^{(\alpha+\beta\epsilon_i)} + 2\gamma}  
    \label{eqn:Order2}
\end{equation}

Using $\gamma = \pm 1/2$, traditional FD or BE statistics can be retrieved from Equation~\ref{eqn:Order2}. The form of Equation~\ref{eqn:Order2} is essentially similar to the probability distribution function derived from AP statistics~\cite{Polychronakos_PhysLettB_1996, Polychronakos_NuclPhysB_1996}. One can generate many more probability distribution functions by solving the polynomial in Equation~\ref{eqn:CFES_2} at different $m$ (see supporting information section SI.V). However, the general solution given in Equation~\ref{eqn:CFES_3} is much too complex to understand the properties of $m$, $\gamma$, and their underlying thermodynamic connections. In the following section, it is shown how a permissible range of $m$ and $\gamma$ factors can be determined for generating a realistic probability distribution function using CFES, via logical and algebraic reasoning.

\section{Properties of CFES}
\label{sect:properties}

To confirm that CFES generates a realistic probability distribution function, one has to consider the dependence of $n_i$ on $\epsilon_i$ such that $\bar{n_i} = \{ \bar{n_i} \in \mathbb{R} | \bar{n_i} \geq 0 \}$ throughout the energy spectrum. Using this basic idea, many interesting features of CFES can be deduced by studying and rearranging Equation~\ref{eqn:CFES_2} itself. Replacing $a=\exp(\alpha+\beta\epsilon_i)/m\gamma$ and $b=(1/m\gamma)$, Equation~\ref{eqn:CFES_2} can be recognized as a polynomial of $\bar{n_i}$ as,

\begin{equation}
    \bar{n_i}^{(m-1)}+a\bar{n_i}-b=0
    \label{eqn:CFES_4}
\end{equation}

For a particular $m \geq 2$, this equation has $(m-1)$ roots, including real and imaginary. For CFES to apply to a system obeying thermodynamic principles, at least one of the root of the Equation~\ref{eqn:CFES_4} must be real and positive. Looking at Equation~\ref{eqn:CFES_4}, it can be found that the two coefficients of the polynomial, $(+a)$ and $(-b)$, have opposite signs at all values of $m$ and $\gamma$. This indicates that at least one real root of $\bar{n_i}$ must exist for all $m > 0$. Using Descartes' sign rule, it is found that there is only one positive root is present at all permissible values of $m$ and $\gamma$. This rules out any ambiguity in the general solution of CFES, and shows that $\bar{n_i}$ always has one, and only one, real positive value at all $\epsilon_i$ for $m > 0$.

\subsection{Limits of m, $\gamma$}

Further rearranging Equation~\ref{eqn:CFES_4}, one can show that, if $\bar{n_i} = \{ \bar{n_i} \in \mathbb{R} | \bar{n_i} \geq 0 \} $, $m \neq 1$, $\gamma \neq 0$, and $m\gamma > 0$; a few possible conditions arise as,

\begin{subequations}
    \begin{align}
        \begin{split}
            &\text{Condition 1: for }  m\gamma > 0 \\
            &\bar{n_i} \leq e^{-(\alpha+\beta\epsilon_i)} 
        \end{split}\\
        \begin{split}
            &\text{Condition 2: for } 0 > m > 1 \text{ and } \gamma > 0 \\
            &\bar{n_i} \geq \Big (\frac{1}{m\gamma} \Big )^{\Big (\frac{1}{m-1} \Big )} 
        \end{split}\\    
        \begin{split}
            &\text{Condition 3: for } m > 1 \text{ and } \gamma > 0 \\
            &\bar{n_i} \leq \Big (\frac{1}{m\gamma} \Big )^{\Big (\frac{1}{m-1} \Big )} \\
        \end{split}\\
        \begin{split}
            &\text{Condition 4: for } m < 0 \text{ and } \gamma < 0 \\  
            &\bar{n_i} \geq \Big (\frac{1}{m\gamma} \Big )^{\Big (\frac{1}{-|m|-1} \Big )} 
        \end{split}
    \end{align} 
    \label{eqn:limit}
\end{subequations}

The condition in Equation~\ref{eqn:limit}a is straightforward, which states that the number of particles at an energy level $\epsilon_i$ will decrease with increasing energy. However, in the rest of the sub-equations in Equation~\ref{eqn:limit}, no energy terms are present. Hence, conditions in Equation~\ref{eqn:limit}b and~\ref{eqn:limit}d are in contradiction with~\ref{eqn:limit}a, as they predict a non-zero lower bound for $\bar{n_i}$, which is independent of energy. This is impossible at $\epsilon_i \to \infty$, where Equation~\ref{eqn:limit}a predicts $\bar{n_i} \to 0$. Hence, $m < 1$ values are not allowed in Equation~\ref{eqn:limit} within the given range of parameters. The third condition in Equation~\ref{eqn:limit}c predicts a constant upper bound, $n_c$, of the occupancy numbers at all available energy levels as, 

\begin{equation}
    n_c = \Bigg({\frac{1}{m\gamma}}\Bigg)^{\Big( \frac{1}{m-1} \Big)}
    \label{eqn:upperbound}
\end{equation}

which should be a valid condition at all $m$ and $\gamma$ values. However, by definition, an \textit{available} energy level means the allowed number of particles cannot be less than one at any energy level, i.e. $n_i^{\text{min}} = 1$. At the limiting case, it means $n_c^{\text{\text{min}}} = 1/g^{\text{\text{min}}}$, where $g^{\text{\text{min}}}$ is lowest available degeneracy at any of the available energy level. Hence, Equation~\ref{eqn:upperbound} can be written as,

\begin{equation}
  \gamma \leq \frac{(g^{\text{\text{min}}})^{(m-1)}}{m}
  \label{eqn:limit_gamma}
\end{equation}

In most classical systems, the degeneracy factor has a monotonous dependence on energy. In fact, most realistic systems (classical or quantum) have a ground state with single degeneracy, i.e. $g^{\text{\text{min}}} = 1$. Hence, as a special case of Equation~\ref{eqn:limit}c and \ref{eqn:limit_gamma} where $g^{\text{\text{min}}} = 1$ and $\gamma > 0$ (labeled as '$r$' and '$+$', for 'real' and 'positive', respectively), one can write,

\begin{equation}
    \gamma_{r}^+ \leq \frac{1}{m}
    \label{eqn:limit_gamma_positive}
\end{equation}

At $m=2$, Equation~\ref{eqn:limit_gamma_positive} yields the FD distribution at the upper limit of $\gamma_r^+$. Hence, all systems which have positive $\gamma$ values and obeys CFES model can be termed as \textit{fermion-like} systems.

Similarly, one can write the possible conditions for $\bar{n_i} = \{ \bar{n_i} \in \mathbb{R} | \bar{n_i} \geq 0 \} $, $m \neq 1$, $\gamma \neq 0$, and $m\gamma < 0$ from Equation~\ref{eqn:CFES_4} as,

\begin{subequations}
    \begin{align}
        \begin{split}
        &\text{Condition 1: for }  m\gamma < 0 \\
        &\bar{n_i} \geq e^{-(\alpha+\beta\epsilon_i)}\\
        \end{split}\\
        \begin{split}
        &\text{Condition 2: for } m > 0 \text{ and } \gamma < 0 \\
        &\bar{n_i} \geq (-1)^{\Big (\frac{1}{m-1} \Big )} \Big (\frac{1}{m|\gamma|} \Big )^{\Big(\frac{1}{m-1}\Big)}\\
        \end{split}\\
        \begin{split}
        &\text{Condition 3: for } m < 0 \text{ and } \gamma > 0 \\
        &\bar{n_i} \geq (-1)^{\Big (\frac{1}{-|m|-1} \Big)} \Big (\frac{1}{|m|\gamma} \Big )^{\Big(\frac{1}{-|m|-1}\Big)}\\ 
        \end{split}
    \end{align}
    \label{eqn:limit_2}                 
\end{subequations}

From Equation~\ref{eqn:limit_2}a, it is clear that $\bar{n_i}$ have a non-zero lower bound at the ground state energy $\epsilon_i = 0$ as $\bar{n_0}^{\text{min}} = \exp(-\alpha)$. As $\bar{n_i}$ is a real number, both sides of all sub-equations in~\ref{eqn:limit_2} has to be real. Using this concept, one can find that in the Equation~\ref{eqn:limit_2}b, negative values of $\gamma$ with $m > 0$ are allowed only for $m = 2$ case. As a result, $n_c$ in Equation~\ref{eqn:limit_2}b becomes negative, which indicates that there are no upper limit in the occupation numbers and FEP does not apply here. From Equation~\ref{eqn:limit_2}c, one can find that negative values of $m$ are not possible to find in a realistic probability distribution function with $\gamma > 0$; as they will produce imaginary numbers. Combining all sub-equations from Equation~\ref{eqn:limit} and~\ref{eqn:limit_2}, one can state that negative or fractional ($< 1$) values of $m$ are not allowed in CFES. 

Performing the same analysis as in Equation~\ref{eqn:limit_gamma} at $m = 2$ and $ \gamma < 0$ (labeled as '$r$' and '$-$', for 'real' and 'negative', respectively) case, one can write,

\begin{equation}
    \gamma_r^- \geq -\frac{1}{m}
    \label{eqn:limit_gamma_negative}
\end{equation}

\begin{figure}[h]
    \centering
    \includegraphics[scale=0.4]{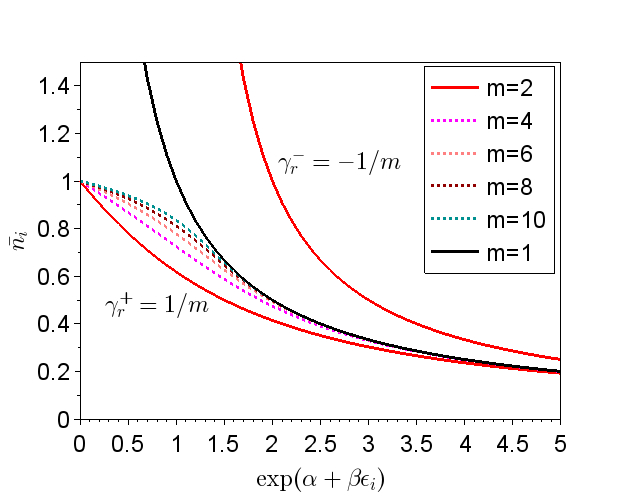}
    \caption{Population distribution function corresponding to CFES model at various $m$. $\gamma_r^+ = 1/m$ and $\gamma_r^- = -1/m$. At m =1, system follows MB statistics and at m=2 system follows FD and BE statistics. Data is produced using Scilab-6.1.1~\cite{scilab} polynomial solver.}
    \label{fig:m_statistics}
\end{figure}

Equation~\ref{eqn:limit_gamma_negative} can be related to the BE statistics at $m = 2$ at the lower bound of $\gamma_r^-$. Therefore, any system that has a negative $\gamma$ value and obeys CFES can be termed as \textit{boson-like} systems. Boson-like systems are quite different from fermion-like systems, as all boson-like systems have a single $m$ value, whereas fermion-like systems have $\infty \geq m \geq 2$. 

In this section, it is shown how the properties of both FD and BE statistics can be generated from the CFES model. From Equation~\ref{eqn:limit_gamma_positive} and Equation~\ref{eqn:limit_gamma_negative}, one can state that for any real systems obeying CFES model with $g^{\text{min}} = 1$, FD and BE statistics are the two limits of the particle distribution. This statement can be clearly understood by observing Figure~\ref{fig:m_statistics}. The choice of $m$ determines how closely the system follows the MB statistics at low energy. From Equation~\ref{eqn:limit_gamma_positive} and ~\ref{eqn:upperbound}, it is evident that at large values of $m \to \infty$, $\gamma_r^+ \sim 0$ and $n_c \to 1$. As $n_i << g_i$, Equation~\ref{eqn:CFES_2} will be reduced to the traditional MB distribution at $m \to \infty$.

The interesting outcome of the CFES is the presence of the upper bound in $\bar{n}_i$ for fermion-like systems, which was not an initial criterion for the derivation in section~\ref{sect:cfes}. In the CFES model, the use of the MaxEnt method leads to a realistic probability function that follows FEP without considering any arguments from the quantum mechanics, other than the concept of DI in the form of a constraint in Equation~\ref{eqn:conditions_1}. Drawing parallels with HW or AP statistics~\cite{Wu_PhysRevLett_1994,Polychronakos_PhysLettB_1996, Polychronakos_NuclPhysB_1996}, one can compare $(1/n_c)$ as the single level statistical interaction parameter defined by Haldane~\cite{Haldane_PhysRevLett_1991} in HW statistics. It was shown earlier~\cite{Bhaduri_Sen_JPhysA_2010} that at infinite statistical correlation among the energy levels, HW statistics becomes analogous to the classical systems. This is similar to the result we have obtained in the CFES model as well. In this paper, however, no {\it mutual} statistics are considered; i.e., no explicit statistical interactions between the populations of two energy levels have been assumed {\it a priori} to derive CFES. The exclusion at each energy level arises naturally to maintain the constrain in Equation~\ref{eqn:conditions_1}. Of course, unlike quantum systems, the applicability of the FEP is valid in CFES only at very dilute conditions.
 
As the permissible ranges of (m,$\gamma$) are now determined, one can now comment on the properties of the weights in Equation~\ref{eqn:CFES_3} and~\ref{eqn:CFES_weight}. Boson-like systems are clearly free from any possibility of negative weights as $\gamma < 0$ and $m > 0$. For fermion-like systems at $m > 0$ and $\gamma > 0$, the positive-negative combinations of the weights cancel out to yield a net positive probability because of the following reasons -- $(a)$ As $m > 0$, the probability factor $\exp[-(\alpha+\beta\epsilon_i)(jm-j+1)]$ is strictly decreasing at $j \to \infty$, $(b)$ $C_j(m)$ is always positive, $(c)$ $\gamma_r^+ \leq (1/m)$, which means the second weight in Equation~\ref{eqn:CFES_weight}, $(-m\gamma)^j \to 0$, at $j \to \infty$. As a result, $\bar{n_i}$ in Equation~\ref{eqn:CFES_3} is always a converging positive function with ($j=even$) sums being larger than ($j=odd$) sums within the permissible range of $m$, $\gamma$. $(d)$ The convergence of the probability function is faster than the HW statistics~\cite{Murthy_Shankar_PhysRevB_1999} due to the presence of the second weight $(m\gamma)^j$, which is fractional within the permissible range. 

\subsection{Scaled probability distribution}

The interpretation of $S_m$ is an important aspect in the CFES model. Although the constraint in Equation~\ref{eqn:conditions_1} may seem empirical, it's use in CFES model generates the probability distribution function corresponding to all three fundamental statistical models. This raises the question, whether this constraint has any physical meaning. In the earlier section, it is shown that this constraint is related to the DI. To relate $S_m$ to thermodynamic quantities, we first rewrite the final equation of CFES in terms of a {\it scaled} particle distribution as,

\begin{equation}
        p_i^{\text{scl}}(\epsilon_i, \beta) = \frac{(p_i^2/g_i) e^{\beta\epsilon_i}}{\sum\limits_i (p_i^2/g_i) e^{\beta\epsilon_i}}
    \label{eqn:P_scl}
\end{equation}

where $p_i^{\text{scl}}$ is the probability of finding a scaled number of particles, $n_i^{\text{scl}} = n_i - \gamma m (n_i^m/g_i^{(m-1)})$, at a certain energy level $\epsilon_i$; and $p_i = n_i/N$ is the actual probability distribution of the CFES model. Details of the derivation is given in supporting information section SI.VI. Equation~\ref{eqn:P_scl} is an alternate version of CFES equation which can be studied without the explicit knowledge of the $m$, $\gamma$ factors. Using the idea of the scaled particle distribution, one can derive the value of $S_m$ using Equation~\ref{eqn:conditions_1}, Equation~\ref{eqn:CFES_2}, and Equation~\ref{eqn:P_scl} as (see supporting information section SI.VI),

\begin{equation}
    S_m = \frac{1}{m\gamma} ( N - N^{\text{scl}} )
    \label{eqn:Sm}
\end{equation}

where $N^{\text{scl}} = \sum_i n_i^{\text{scl}}$ is the total scaled particle number. $S_m$ thus represents the total {\it effective} particle number, $N^{\text{eff}} = N - N^{\text{scl}}$, which is scaled by $m\gamma$ factor. $S_m$, $m$, and $\gamma$ are independent quantities but together they determine $N^{\text{eff}}$.  

\section{Application of CFES to classical thermodynamics }
\label{sect:thermodynamics}

The relationship between any statistical model and thermodynamics depends on the equation of microstate counting. Using various forms microstate counting formula, the application of QFES models in quantum thermodynamics has been studied extensively in the literature~\cite{Anghel_JPhysConfSer_2013}. However, in this paper, the CFES model is derived from a classical framework. Naturally, the application of CFES in classical thermodynamics is a main focus in this section. 

\subsection{Free Energy equation}

Ideally, the Gibbs equation of chemical potential for classical particles at equilibrium ($\mu(\beta, m, \gamma)$) can be written as,

\begin{equation}
    N\mu = E + \frac{N}{\beta} - \frac{\ln(\widetilde{W})}{\beta} - \frac{\chi}{\beta}
    \label{eqn:gibbs}
\end{equation}

where $\widetilde{S}=\ln(\widetilde{W})$ is the total entropy of the system, and $\chi/\beta$  -- namely the {\it exclusion potential} -- represents the excess contribution to the free energy arising due to the dependence of $\mu$ on the $m$ and $\gamma$ factors. Equation~\ref{eqn:gibbs} is akin to the thermodynamic equation used in the field of nanothermodynamics~\cite{Hill_JChemPhys_1962, Hill_NanoLett_2001, Hill_NanoLett_2001_1, Hill_NanoLett_2001_2, Chamberlin_Entropy_2015, Chamberlin_Wolf_Symmetry_2021}, where a {\it subdivision potential} was used to describe the thermodynamics of a subsystem. However, origin of the exclusion potential doesn't lie on the segmentation of the system, rather on the FEP maintained at each energy level. For the CFES model, it is straightforward to show that, $\alpha = -\mu\beta$. However, definition of $\beta$ is reserved for later. In equilibrium, $\chi(\beta, m, \gamma)$ has the following form,

\begin{equation}
    \chi = \sum_i n_i \ln \Big(1 - m\gamma \bar{n}_i^{(m-1)} \Big )
    \label{eqn:chi}
\end{equation}

The constant $\chi$ factor controls the occupation constraints discussed in the previous section as $\{ \chi \in \mathbb{R} \}$. For fermion-like systems, $\chi$ is negative, and for boson-like systems, $\chi$ is positive. $\chi$ is a function of both the distribution of the particles in the energy levels, as well as the degeneracy factor $g_i$. Using the concept of the scaled probability distribution, one can simplify $\chi$ (see supporting information section SI.VII) as,

\begin{equation}
        \chi = \ln\widetilde{W} - \ln\widetilde{W}^{\text{scl}}
\label{eqn:chi_2}
\end{equation}

where, $\widetilde{S}^{\text{scl}}=\ln\widetilde{W}^{\text{scl}}$ is the cross-entropic term where $\widetilde{W}^{\text{scl}}$ is defined as,

\begin{equation}
    \widetilde{W}^{\text{scl}} = \prod_i \Bigg(\frac{g_i^{n_i}}{n_i!} \Bigg) \Bigg(\frac{n_i}{n_i^{\text{scl}}} \Bigg)^{n_i}
    \label{eqn:microstates_red}
\end{equation}

Equation~\ref{eqn:chi_2} shows that, $\chi$ is related to entropy and $S^{\text{scl}}=\ln \widetilde{W}^{\text{scl}}$ is essentially a correction to the total entropy of the system. Inserting Equation~\ref{eqn:chi_2} in Equation~\ref{eqn:gibbs}, we get,

\begin{equation}
    N\mu = E + \frac{N}{\beta} - \frac{\widetilde{S}^{\text{eff}}}{\beta}
    \label{eqn:gibbs_1}
\end{equation}    

where $\widetilde{S}^{\text{eff}} = 2\widetilde{S} - \widetilde{S}^{\text{scl}}$ the effective entropy of the system. From Equation~\ref{eqn:microstates_red}, it is evident that if $\gamma$ is positive (i.e. fermion-like systems), $\widetilde{S}^{\text{scl}} > \widetilde{S}$ and subsequently $\widetilde{S}^{\text{eff}}$ will decrease as compared to $\widetilde{S}$, and vice verse. At constant volume $v$, inverse temperature of the system is defined as $\beta \equiv (\partial \widetilde{S}^{\text{eff}}/ \partial E)_v$. Interestingly, $\beta \neq \beta_{\text{MB}}$, where $\beta_{\text{MB}}$ corresponds to the inverse temperature derived form the definition in traditional MB statistics, i.e., $\beta_{\text{MB}} \equiv (\partial S/\partial E)_v$. Therefore, the definition of temperature in CFES model is distinct from MB statistics. One can show that at a concentrated system ($n_i \sim g_i$ or $n_i >> g_i$, where Equation~\ref{eqn:CFES_1} cannot be approximated to Equation~\ref{eqn:CFES_2}) $\beta = \beta_{\text{MB}}$. The distinction arises because $\chi$ derived from Equation~\ref{eqn:CFES_2} (dilute case) is directly related to the entropy. This results in, $\chi \equiv f(E)$ and $(\partial \chi/\partial E)_v \neq 0$. Whereas, $\chi$ derived from Equation~\ref{eqn:CFES_1} (concentrated case) will be independent of entropy or energy as,

\begin{equation}
	\chi_{\text{conc.}} = m\gamma S_m = constant
	\label{eqn:chi_conc}
\end{equation} 

\subsection{Effective temperature approximation}

The equivalence between $\chi$ and system energy allows simplifying CFES to a classical ideal gas scenario, which obeys slightly modified MB statistics. If the particles are not interacting with one another, the entropy of the system has a direct proportionality to the total energy of the system. Assuming that $n_i \sim n_i^{\text{scl}}$ -- i.e. $\chi$ is relatively small -- one can approximate the exclusion potential as,

\begin{equation}
    \frac{\chi}{\beta} \approx \lambda E
    \label{eqn:lambda}
\end{equation}

where $\lambda \equiv f(\beta, m, \gamma)$ is a constant and independent of energy. Substituting Equation~\ref{eqn:lambda} to Equation~\ref{eqn:gibbs}, an effective energy parameter can be found as, $E^{\text{eff}} = E(1-\lambda)$. Using this effective energy in place of the system energy, the usual Gibbs free energy equation is recovered, where $\beta(1 - \lambda) = (\partial S/\partial E)_v$. Here, an effective inverse temperature, $\beta^{\text{eff}} = \beta(1-\lambda)$, replaces the original inverse temperature ($\beta$) of the system. The advantage of reformulating Equation~\ref{eqn:gibbs} with effective energy is that, one can easily derive the corresponding MB statistics in a retrospective way as,

\begin{equation}
    p_i \sim g_ie^{-\beta(1-\lambda)\epsilon_i}
    \label{eqn:MB_effective}
\end{equation}

Equation~\ref{eqn:MB_effective} is valid only if $\lambda$ is not a function of energy. This description of CFES is essentially the same as the ideal gas scenario in MB statistics, albeit with an effective temperature. For fermion-like systems, $n_i^{\text{scl}} < n_i$ from Equation~\ref{eqn:P_scl}; hence $\widetilde{W}^{\text{scl}} > \widetilde{W}$ and $\chi,\lambda < 0$. Hence, from the effective inverse temperature relation, $\beta^{\text{eff}} = \beta(1-\lambda)$, one can state that the effective temperature in fermion-like systems will be lower than the system temperature. This means, the presence of FEP in an ideal gas will decrease the effective system temperature as compared to the original temperature, as part of the entropy is lost to maintain the FEP. Similarly, for boson-like systems, $\chi,\lambda > 0$ and $\beta^{\text{eff}} < \beta$. 

\section{Conclusion and Outlook}

In this paper, the classical analog of the FES model is investigated and analyzed. We have shown that an additive DI parameter can be isolated from the BE microstate formula, which linearly combine with $S_{\text{MB}}$ to generate $S_{\text{BE}}$. Such a linear relationship between $S_{\text{BE}}$ and $S_{\text{MB}}$ is interesting, as it suggests that BE statistics can be generated from MB statistics via MaxEnt methods. The CFES model is derived based on this idea. Moreover, it is shown that CFES model predicts presence of FEP in fermion-like systems. However, it should be noted that such quantum effects are only applicable in the CFES model when the system is sufficiently dilute. Since we have considered a classical system, the microstate counting formula in CFES is not changed from the MB statistics. Given the similarity of CFES with AP statistics, it is perhaps possible to derive a microstate counting equation that can generate similar parental equations as CFES for hypothetical quantum systems. In this way, the current version of the CFES model may have a quantum interpretation as well. 

The constraint in Equation~\ref{eqn:conditions_1} is a semi-empirical one, as it is related to the DI. The presence of DI in classical systems can be explained if some of the degenerate particles of the same species can be distinguished by additional degrees of freedom, which are not incorporated in the statistics model~\cite{Medvedev_PhysRevLett_1997}. However, presence of these additional degrees of freedom allows the degeneracy of the energy levels to vary which, in turn, affect the entropy of the system. If such degrees of freedom are measurable and incorporated in the model, then DI becomes the traditional binary function. However, if such quantities are not measurable, then one has to consider that DI is a variable quantity. More theoretical work is needed to understand the concept of DI in classical systems.

The microcanonical CFES model shown in this paper should be easily extended to Canonical systems and grand-Canonical systems as well. Derivation of several thermodynamic quantities, such as specific heat capacities, virial coefficients, thermodynamic potential etc., and their relation to the partition function $Z_m$ is the next logical step to understand the relation between thermodynamics and different parameters in CFES. The current version of the CFES model assumes particles in an ideal gas scenario with no interactions among each other. Further mathematical analysis is needed to understand the effect of pairwise interactions~\cite{Anghel_PhysLettA_2008, Anghel_PhysScr_2012} in a CFES model. Of course, the present form of CFES is valid for a uniform maximum occupancy at all energy levels. One needs to investigate the properties of CFES in case different energy levels have different $\bar{n_c}$.  

It will be interesting to find realistic classical systems where the CFES model is applicable. One possible candidate is the two-dimensional silica (2D-silica)~\cite{Lichtenstein_Freund_AngewChem_2012}. Previously, in the equilibrium simulation of two-dimensional silica melt using molecular dynamics~\cite{Roy_Heuer_PhysChemChemPhys_2018}, the presence of effective temperature~\cite{Roy_Heuer_PhysRevLett_2019, Roy_Heuer_JPhysCondMat_2019} was observed in the probability distribution of various ring-sizes. It was found that the probability distributions of the individual rings also show the presence of an effective temperature. Here, the deviation in the temperature factor was attributed to the fact that the rings are spatially correlated in such a way, that there are certain preferences among the neighbours~\cite{Roy_Heuer_PhysRevLett_2019}. Due to these correlations, the density of states of the individual rings are reduced, and the statistics of the system deviates from the supposed MB statistics. It will be interesting to apply the CFES technique in 2D-silica to understand the presence of effective temperature in terms of FEP, and quantify the extent of correlations in terms of $m$, $\gamma$ parameters. 

\section*{Acknowledgement}

The author would like to thank Prof. Pinaki Chaudhuri, Prof. M.V.N. Murthy, Prof. R. Shankar, and Umang Dattani at IMSc., Chennai for useful discussions.


\begin{thebibliography}{76}%
\makeatletter
\providecommand \@ifxundefined [1]{%
 \@ifx{#1\undefined}
}%
\providecommand \@ifnum [1]{%
 \ifnum #1\expandafter \@firstoftwo
 \else \expandafter \@secondoftwo
 \fi
}%
\providecommand \@ifx [1]{%
 \ifx #1\expandafter \@firstoftwo
 \else \expandafter \@secondoftwo
 \fi
}%
\providecommand \natexlab [1]{#1}%
\providecommand \enquote  [1]{``#1''}%
\providecommand \bibnamefont  [1]{#1}%
\providecommand \bibfnamefont [1]{#1}%
\providecommand \citenamefont [1]{#1}%
\providecommand \href@noop [0]{\@secondoftwo}%
\providecommand \href [0]{\begingroup \@sanitize@url \@href}%
\providecommand \@href[1]{\@@startlink{#1}\@@href}%
\providecommand \@@href[1]{\endgroup#1\@@endlink}%
\providecommand \@sanitize@url [0]{\catcode `\\12\catcode `\$12\catcode
  `\&12\catcode `\#12\catcode `\^12\catcode `\_12\catcode `\%12\relax}%
\providecommand \@@startlink[1]{}%
\providecommand \@@endlink[0]{}%
\providecommand \url  [0]{\begingroup\@sanitize@url \@url }%
\providecommand \@url [1]{\endgroup\@href {#1}{\urlprefix }}%
\providecommand \urlprefix  [0]{URL }%
\providecommand \Eprint [0]{\href }%
\providecommand \doibase [0]{http://dx.doi.org/}%
\providecommand \selectlanguage [0]{\@gobble}%
\providecommand \bibinfo  [0]{\@secondoftwo}%
\providecommand \bibfield  [0]{\@secondoftwo}%
\providecommand \translation [1]{[#1]}%
\providecommand \BibitemOpen [0]{}%
\providecommand \bibitemStop [0]{}%
\providecommand \bibitemNoStop [0]{.\EOS\space}%
\providecommand \EOS [0]{\spacefactor3000\relax}%
\providecommand \BibitemShut  [1]{\csname bibitem#1\endcsname}%
\let\auto@bib@innerbib\@empty
\bibitem [{\citenamefont {Milotti}\ \emph {et~al.}(2007)\citenamefont
  {Milotti}, \citenamefont {Bartalucci}, \citenamefont {Bertolucci},
  \citenamefont {Bragadireanu}, \citenamefont {Cargnelli}, \citenamefont
  {Catitti}, \citenamefont {Curceanu~(Petrascu)}, \citenamefont {Di~Matteo},
  \citenamefont {Egger}, \citenamefont {Guaraldo}, \citenamefont {Iliescu},
  \citenamefont {Ishiwatari}, \citenamefont {Laubenstein}, \citenamefont
  {Marton}, \citenamefont {Pietreanu}, \citenamefont {Ponta}, \citenamefont
  {Sirghi}, \citenamefont {Sirghi}, \citenamefont {Sperandio}, \citenamefont
  {Widmann},\ and\ \citenamefont
  {Zmeskal}}]{Milotti_Zmeskal_IntJModPhysA_2007}%
  \BibitemOpen
  \bibfield  {author} {\bibinfo {author} {\bibfnamefont {E.}~\bibnamefont
  {Milotti}}, \bibinfo {author} {\bibfnamefont {S.}~\bibnamefont {Bartalucci}},
  \bibinfo {author} {\bibfnamefont {S.}~\bibnamefont {Bertolucci}}, \bibinfo
  {author} {\bibfnamefont {M.}~\bibnamefont {Bragadireanu}}, \bibinfo {author}
  {\bibfnamefont {M.}~\bibnamefont {Cargnelli}}, \bibinfo {author}
  {\bibfnamefont {M.}~\bibnamefont {Catitti}}, \bibinfo {author} {\bibfnamefont
  {C.}~\bibnamefont {Curceanu~(Petrascu)}}, \bibinfo {author} {\bibfnamefont
  {S.}~\bibnamefont {Di~Matteo}}, \bibinfo {author} {\bibfnamefont {J.-P.}\
  \bibnamefont {Egger}}, \bibinfo {author} {\bibfnamefont {C.}~\bibnamefont
  {Guaraldo}}, \bibinfo {author} {\bibfnamefont {M.}~\bibnamefont {Iliescu}},
  \bibinfo {author} {\bibfnamefont {T.}~\bibnamefont {Ishiwatari}}, \bibinfo
  {author} {\bibfnamefont {M.}~\bibnamefont {Laubenstein}}, \bibinfo {author}
  {\bibfnamefont {J.}~\bibnamefont {Marton}}, \bibinfo {author} {\bibfnamefont
  {D.}~\bibnamefont {Pietreanu}}, \bibinfo {author} {\bibfnamefont
  {T.}~\bibnamefont {Ponta}}, \bibinfo {author} {\bibfnamefont {D.~L.}\
  \bibnamefont {Sirghi}}, \bibinfo {author} {\bibfnamefont {F.}~\bibnamefont
  {Sirghi}}, \bibinfo {author} {\bibfnamefont {L.}~\bibnamefont {Sperandio}},
  \bibinfo {author} {\bibfnamefont {E.}~\bibnamefont {Widmann}}, \ and\
  \bibinfo {author} {\bibfnamefont {J.}~\bibnamefont {Zmeskal}},\ }\href
  {\doibase 10.1142/S0217751X07035392} {\bibfield  {journal} {\bibinfo
  {journal} {International Journal of Modern Physics A}\ }\textbf {\bibinfo
  {volume} {22}},\ \bibinfo {pages} {242} (\bibinfo {year} {2007})}\BibitemShut
  {NoStop}%
\bibitem [{\citenamefont {Gentile~j.}(1940)}]{Gentile_NuovoCimento_1940}%
  \BibitemOpen
  \bibfield  {author} {\bibinfo {author} {\bibfnamefont {G.}~\bibnamefont
  {Gentile~j.}},\ }\href {\doibase 10.1007/BF02960187} {\bibfield  {journal}
  {\bibinfo  {journal} {Il Nuovo Cimento (1924-1942)}\ }\textbf {\bibinfo
  {volume} {17}},\ \bibinfo {pages} {493} (\bibinfo {year} {1940})}\BibitemShut
  {NoStop}%
\bibitem [{\citenamefont {Tsallis}(1988)}]{Tsallis_JStatPhys_1988}%
  \BibitemOpen
  \bibfield  {author} {\bibinfo {author} {\bibfnamefont {C.}~\bibnamefont
  {Tsallis}},\ }\href {\doibase 10.1007/BF01016429} {\bibfield  {journal}
  {\bibinfo  {journal} {Journal of Statistical Physics}\ }\textbf {\bibinfo
  {volume} {52}},\ \bibinfo {pages} {479} (\bibinfo {year} {1988})}\BibitemShut
  {NoStop}%
\bibitem [{\citenamefont {March}\ \emph {et~al.}(1993)\citenamefont {March},
  \citenamefont {Gidopoulos}, \citenamefont {Theophilou}, \citenamefont {Lea},\
  and\ \citenamefont {Sung}}]{March_Sung_PhysChemLiq_1993}%
  \BibitemOpen
  \bibfield  {author} {\bibinfo {author} {\bibfnamefont {N.~H.}\ \bibnamefont
  {March}}, \bibinfo {author} {\bibfnamefont {N.}~\bibnamefont {Gidopoulos}},
  \bibinfo {author} {\bibfnamefont {A.~K.}\ \bibnamefont {Theophilou}},
  \bibinfo {author} {\bibfnamefont {M.~J.}\ \bibnamefont {Lea}}, \ and\
  \bibinfo {author} {\bibfnamefont {W.}~\bibnamefont {Sung}},\ }\href {\doibase
  10.1080/00319109308030827} {\bibfield  {journal} {\bibinfo  {journal}
  {Physics and Chemistry of Liquids}\ }\textbf {\bibinfo {volume} {26}},\
  \bibinfo {pages} {135} (\bibinfo {year} {1993})}\BibitemShut {NoStop}%
\bibitem [{\citenamefont {March}(1997)}]{March_PhysChemLiq_1997}%
  \BibitemOpen
  \bibfield  {author} {\bibinfo {author} {\bibfnamefont {N.~H.}\ \bibnamefont
  {March}},\ }\href {\doibase 10.1080/00319109708035914} {\bibfield  {journal}
  {\bibinfo  {journal} {Physics and Chemistry of Liquids}\ }\textbf {\bibinfo
  {volume} {34}},\ \bibinfo {pages} {61} (\bibinfo {year} {1997})}\BibitemShut
  {NoStop}%
\bibitem [{\citenamefont {Acharya}\ and\ \citenamefont
  {Swamy}(1994)}]{Acharya_Swamy_JPhysA_1994}%
  \BibitemOpen
  \bibfield  {author} {\bibinfo {author} {\bibfnamefont {R.}~\bibnamefont
  {Acharya}}\ and\ \bibinfo {author} {\bibfnamefont {P.~N.}\ \bibnamefont
  {Swamy}},\ }\href {\doibase 10.1088/0305-4470/27/22/005} {\bibfield
  {journal} {\bibinfo  {journal} {Journal of Physics A: Mathematical and
  General}\ }\textbf {\bibinfo {volume} {27}},\ \bibinfo {pages} {7247}
  (\bibinfo {year} {1994})}\BibitemShut {NoStop}%
\bibitem [{\citenamefont {Khare}(2005)}]{Khare_2005}%
  \BibitemOpen
  \bibfield  {author} {\bibinfo {author} {\bibfnamefont {A.}~\bibnamefont
  {Khare}},\ }\href@noop {} {\emph {\bibinfo {title} {Fractional statistics and
  quantum theory}}}\ (\bibinfo  {publisher} {World Scientific},\ \bibinfo
  {year} {2005})\BibitemShut {NoStop}%
\bibitem [{\citenamefont {Nayak}\ \emph {et~al.}(2008)\citenamefont {Nayak},
  \citenamefont {Simon}, \citenamefont {Stern}, \citenamefont {Freedman},\ and\
  \citenamefont {Das~Sarma}}]{Nayak_Sarma_RevModPhys_2008}%
  \BibitemOpen
  \bibfield  {author} {\bibinfo {author} {\bibfnamefont {C.}~\bibnamefont
  {Nayak}}, \bibinfo {author} {\bibfnamefont {S.~H.}\ \bibnamefont {Simon}},
  \bibinfo {author} {\bibfnamefont {A.}~\bibnamefont {Stern}}, \bibinfo
  {author} {\bibfnamefont {M.}~\bibnamefont {Freedman}}, \ and\ \bibinfo
  {author} {\bibfnamefont {S.}~\bibnamefont {Das~Sarma}},\ }\href {\doibase
  10.1103/RevModPhys.80.1083} {\bibfield  {journal} {\bibinfo  {journal} {Rev.
  Mod. Phys.}\ }\textbf {\bibinfo {volume} {80}},\ \bibinfo {pages} {1083}
  (\bibinfo {year} {2008})}\BibitemShut {NoStop}%
\bibitem [{\citenamefont {Dasni\`eres~de Veigy}\ and\ \citenamefont
  {Ouvry}(1994)}]{Veigy_Ovury_PhysRevLett_1994}%
  \BibitemOpen
  \bibfield  {author} {\bibinfo {author} {\bibfnamefont {A.}~\bibnamefont
  {Dasni\`eres~de Veigy}}\ and\ \bibinfo {author} {\bibfnamefont
  {S.}~\bibnamefont {Ouvry}},\ }\href {\doibase 10.1103/PhysRevLett.72.600}
  {\bibfield  {journal} {\bibinfo  {journal} {Phys. Rev. Lett.}\ }\textbf
  {\bibinfo {volume} {72}},\ \bibinfo {pages} {600} (\bibinfo {year}
  {1994})}\BibitemShut {NoStop}%
\bibitem [{\citenamefont {Dasni\`eres~de Veigy}\ and\ \citenamefont
  {Ouvry}(1995)}]{Veigy_Ovury_ModPhysLettA_1995_1}%
  \BibitemOpen
  \bibfield  {author} {\bibinfo {author} {\bibfnamefont {A.}~\bibnamefont
  {Dasni\`eres~de Veigy}}\ and\ \bibinfo {author} {\bibfnamefont
  {S.}~\bibnamefont {Ouvry}},\ }\href {\doibase 10.1142/S0217732395000028}
  {\bibfield  {journal} {\bibinfo  {journal} {Modern Physics Letters A}\
  }\textbf {\bibinfo {volume} {10}},\ \bibinfo {pages} {1} (\bibinfo {year}
  {1995})}\BibitemShut {NoStop}%
\bibitem [{\citenamefont {de~Veigy}\ and\ \citenamefont
  {Ouvry}(1995)}]{Veigy_Ovury_ModPhysLettA_1995_2}%
  \BibitemOpen
  \bibfield  {author} {\bibinfo {author} {\bibfnamefont {A.~D.}\ \bibnamefont
  {de~Veigy}}\ and\ \bibinfo {author} {\bibfnamefont {S.}~\bibnamefont
  {Ouvry}},\ }\href {\doibase 10.1142/S0217984995000267} {\bibfield  {journal}
  {\bibinfo  {journal} {Modern Physics Letters B}\ }\textbf {\bibinfo {volume}
  {09}},\ \bibinfo {pages} {271} (\bibinfo {year} {1995})}\BibitemShut
  {NoStop}%
\bibitem [{\citenamefont {Bartolomei}\ \emph {et~al.}(2020)\citenamefont
  {Bartolomei}, \citenamefont {Kumar}, \citenamefont {Bisognin}, \citenamefont
  {Marguerite}, \citenamefont {Berroir}, \citenamefont {Bocquillon},
  \citenamefont {Plaçais}, \citenamefont {Cavanna}, \citenamefont {Dong},
  \citenamefont {Gennser}, \citenamefont {Jin},\ and\ \citenamefont
  {Fève}}]{Bartolomei_Feve_Science_2020}%
  \BibitemOpen
  \bibfield  {author} {\bibinfo {author} {\bibfnamefont {H.}~\bibnamefont
  {Bartolomei}}, \bibinfo {author} {\bibfnamefont {M.}~\bibnamefont {Kumar}},
  \bibinfo {author} {\bibfnamefont {R.}~\bibnamefont {Bisognin}}, \bibinfo
  {author} {\bibfnamefont {A.}~\bibnamefont {Marguerite}}, \bibinfo {author}
  {\bibfnamefont {J.-M.}\ \bibnamefont {Berroir}}, \bibinfo {author}
  {\bibfnamefont {E.}~\bibnamefont {Bocquillon}}, \bibinfo {author}
  {\bibfnamefont {B.}~\bibnamefont {Plaçais}}, \bibinfo {author}
  {\bibfnamefont {A.}~\bibnamefont {Cavanna}}, \bibinfo {author} {\bibfnamefont
  {Q.}~\bibnamefont {Dong}}, \bibinfo {author} {\bibfnamefont {U.}~\bibnamefont
  {Gennser}}, \bibinfo {author} {\bibfnamefont {Y.}~\bibnamefont {Jin}}, \ and\
  \bibinfo {author} {\bibfnamefont {G.}~\bibnamefont {Fève}},\ }\href
  {\doibase 10.1126/science.aaz5601} {\bibfield  {journal} {\bibinfo  {journal}
  {Science}\ }\textbf {\bibinfo {volume} {368}},\ \bibinfo {pages} {173}
  (\bibinfo {year} {2020})}\BibitemShut {NoStop}%
\bibitem [{\citenamefont {Stern}(2008)}]{Stern_AnnPhys_2008}%
  \BibitemOpen
  \bibfield  {author} {\bibinfo {author} {\bibfnamefont {A.}~\bibnamefont
  {Stern}},\ }\href {\doibase 10.1016/j.aop.2007.10.008} {\bibfield  {journal}
  {\bibinfo  {journal} {Annals of Physics}\ }\textbf {\bibinfo {volume}
  {323}},\ \bibinfo {pages} {204} (\bibinfo {year} {2008})},\ \bibinfo {note}
  {january Special Issue 2008}\BibitemShut {NoStop}%
\bibitem [{\citenamefont
  {Laughlin}(1988{\natexlab{a}})}]{Laughlin_PhysRevLett_1988}%
  \BibitemOpen
  \bibfield  {author} {\bibinfo {author} {\bibfnamefont {R.~B.}\ \bibnamefont
  {Laughlin}},\ }\href {\doibase 10.1103/PhysRevLett.60.2677} {\bibfield
  {journal} {\bibinfo  {journal} {Phys. Rev. Lett.}\ }\textbf {\bibinfo
  {volume} {60}},\ \bibinfo {pages} {2677} (\bibinfo {year}
  {1988}{\natexlab{a}})}\BibitemShut {NoStop}%
\bibitem [{\citenamefont
  {Laughlin}(1988{\natexlab{b}})}]{Laughlin_Science_1988}%
  \BibitemOpen
  \bibfield  {author} {\bibinfo {author} {\bibfnamefont {R.~B.}\ \bibnamefont
  {Laughlin}},\ }\href {\doibase 10.1126/science.242.4878.525} {\bibfield
  {journal} {\bibinfo  {journal} {Science}\ }\textbf {\bibinfo {volume}
  {242}},\ \bibinfo {pages} {525} (\bibinfo {year}
  {1988}{\natexlab{b}})}\BibitemShut {NoStop}%
\bibitem [{\citenamefont {{Aghaei Abchouyeh}}\ \emph
  {et~al.}(2018)\citenamefont {{Aghaei Abchouyeh}}, \citenamefont {Mirza},
  \citenamefont {{Karimi Takrami}},\ and\ \citenamefont
  {Younesizadeh}}]{Abchouyeh_Younesizadeh_PhysLettB_2018}%
  \BibitemOpen
  \bibfield  {author} {\bibinfo {author} {\bibfnamefont {M.}~\bibnamefont
  {{Aghaei Abchouyeh}}}, \bibinfo {author} {\bibfnamefont {B.}~\bibnamefont
  {Mirza}}, \bibinfo {author} {\bibfnamefont {M.}~\bibnamefont {{Karimi
  Takrami}}}, \ and\ \bibinfo {author} {\bibfnamefont {Y.}~\bibnamefont
  {Younesizadeh}},\ }\href {\doibase 10.1016/j.physletb.2018.02.066} {\bibfield
   {journal} {\bibinfo  {journal} {Physics Letters B}\ }\textbf {\bibinfo
  {volume} {780}},\ \bibinfo {pages} {240} (\bibinfo {year}
  {2018})}\BibitemShut {NoStop}%
\bibitem [{\citenamefont {Haldane}(1991)}]{Haldane_PhysRevLett_1991}%
  \BibitemOpen
  \bibfield  {author} {\bibinfo {author} {\bibfnamefont {F.~D.~M.}\
  \bibnamefont {Haldane}},\ }\href {\doibase 10.1103/PhysRevLett.67.937}
  {\bibfield  {journal} {\bibinfo  {journal} {Phys. Rev. Lett.}\ }\textbf
  {\bibinfo {volume} {67}},\ \bibinfo {pages} {937} (\bibinfo {year}
  {1991})}\BibitemShut {NoStop}%
\bibitem [{\citenamefont {Wu}(1994)}]{Wu_PhysRevLett_1994}%
  \BibitemOpen
  \bibfield  {author} {\bibinfo {author} {\bibfnamefont {Y.-S.}\ \bibnamefont
  {Wu}},\ }\href {\doibase 10.1103/PhysRevLett.73.922} {\bibfield  {journal}
  {\bibinfo  {journal} {Phys. Rev. Lett.}\ }\textbf {\bibinfo {volume} {73}},\
  \bibinfo {pages} {922} (\bibinfo {year} {1994})}\BibitemShut {NoStop}%
\bibitem [{\citenamefont {Nayak}\ and\ \citenamefont
  {Wilczek}(1994)}]{Nayak_Wilczek_PhysRevLett_1994}%
  \BibitemOpen
  \bibfield  {author} {\bibinfo {author} {\bibfnamefont {C.}~\bibnamefont
  {Nayak}}\ and\ \bibinfo {author} {\bibfnamefont {F.}~\bibnamefont
  {Wilczek}},\ }\href {\doibase 10.1103/PhysRevLett.73.2740} {\bibfield
  {journal} {\bibinfo  {journal} {Phys. Rev. Lett.}\ }\textbf {\bibinfo
  {volume} {73}},\ \bibinfo {pages} {2740} (\bibinfo {year}
  {1994})}\BibitemShut {NoStop}%
\bibitem [{\citenamefont {Anghel}(2007)}]{Anghel_JPhysA_2007}%
  \BibitemOpen
  \bibfield  {author} {\bibinfo {author} {\bibfnamefont {D.-V.}\ \bibnamefont
  {Anghel}},\ }\href {\doibase 10.1088/1751-8113/40/47/f01} {\bibfield
  {journal} {\bibinfo  {journal} {Journal of Physics A: Mathematical and
  Theoretical}\ }\textbf {\bibinfo {volume} {40}},\ \bibinfo {pages} {F1013}
  (\bibinfo {year} {2007})}\BibitemShut {NoStop}%
\bibitem [{\citenamefont {Anghel}(2009)}]{Anghel_EurPhysLett_2009}%
  \BibitemOpen
  \bibfield  {author} {\bibinfo {author} {\bibfnamefont {D.-V.}\ \bibnamefont
  {Anghel}},\ }\href {\doibase 10.1209/0295-5075/87/60009} {\bibfield
  {journal} {\bibinfo  {journal} {{EPL} (Europhysics Letters)}\ }\textbf
  {\bibinfo {volume} {87}},\ \bibinfo {pages} {60009} (\bibinfo {year}
  {2009})}\BibitemShut {NoStop}%
\bibitem [{\citenamefont {Anghel}(2013)}]{Anghel_JPhysConfSer_2013}%
  \BibitemOpen
  \bibfield  {author} {\bibinfo {author} {\bibfnamefont {D.-V.}\ \bibnamefont
  {Anghel}},\ }\href {\doibase 10.1088/1742-6596/410/1/012121} {\bibfield
  {journal} {\bibinfo  {journal} {Journal of Physics: Conference Series}\
  }\textbf {\bibinfo {volume} {410}},\ \bibinfo {pages} {012121} (\bibinfo
  {year} {2013})}\BibitemShut {NoStop}%
\bibitem [{\citenamefont
  {Isakov}(1994{\natexlab{a}})}]{Isakov_PhysRevLett_1994}%
  \BibitemOpen
  \bibfield  {author} {\bibinfo {author} {\bibfnamefont {S.~B.}\ \bibnamefont
  {Isakov}},\ }\href {\doibase 10.1103/PhysRevLett.73.2150} {\bibfield
  {journal} {\bibinfo  {journal} {Phys. Rev. Lett.}\ }\textbf {\bibinfo
  {volume} {73}},\ \bibinfo {pages} {2150} (\bibinfo {year}
  {1994}{\natexlab{a}})}\BibitemShut {NoStop}%
\bibitem [{\citenamefont
  {Isakov}(1994{\natexlab{b}})}]{Isakov_ModPhysLettB_1994}%
  \BibitemOpen
  \bibfield  {author} {\bibinfo {author} {\bibfnamefont {S.~B.}\ \bibnamefont
  {Isakov}},\ }\href {\doibase 10.1142/S0217984994000327} {\bibfield  {journal}
  {\bibinfo  {journal} {Modern Physics Letters B}\ }\textbf {\bibinfo {volume}
  {08}},\ \bibinfo {pages} {319} (\bibinfo {year}
  {1994}{\natexlab{b}})}\BibitemShut {NoStop}%
\bibitem [{\citenamefont {Jaynes}(1957{\natexlab{a}})}]{Jaynes_PhysRev_1957}%
  \BibitemOpen
  \bibfield  {author} {\bibinfo {author} {\bibfnamefont {E.~T.}\ \bibnamefont
  {Jaynes}},\ }\href {\doibase 10.1103/PhysRev.106.620} {\bibfield  {journal}
  {\bibinfo  {journal} {Phys. Rev.}\ }\textbf {\bibinfo {volume} {106}},\
  \bibinfo {pages} {620} (\bibinfo {year} {1957}{\natexlab{a}})}\BibitemShut
  {NoStop}%
\bibitem [{\citenamefont {Jaynes}(1957{\natexlab{b}})}]{Jaynes_PhysRev_1957_2}%
  \BibitemOpen
  \bibfield  {author} {\bibinfo {author} {\bibfnamefont {E.~T.}\ \bibnamefont
  {Jaynes}},\ }\href {\doibase 10.1103/PhysRev.108.171} {\bibfield  {journal}
  {\bibinfo  {journal} {Phys. Rev.}\ }\textbf {\bibinfo {volume} {108}},\
  \bibinfo {pages} {171} (\bibinfo {year} {1957}{\natexlab{b}})}\BibitemShut
  {NoStop}%
\bibitem [{\citenamefont {Ramanathan}(1992)}]{Ramanathan_PhysRevD_1992}%
  \BibitemOpen
  \bibfield  {author} {\bibinfo {author} {\bibfnamefont {R.}~\bibnamefont
  {Ramanathan}},\ }\href {\doibase 10.1103/PhysRevD.45.4706} {\bibfield
  {journal} {\bibinfo  {journal} {Phys. Rev. D}\ }\textbf {\bibinfo {volume}
  {45}},\ \bibinfo {pages} {4706} (\bibinfo {year} {1992})}\BibitemShut
  {NoStop}%
\bibitem [{\citenamefont {Camino}\ \emph {et~al.}(2005)\citenamefont {Camino},
  \citenamefont {Zhou},\ and\ \citenamefont
  {Goldman}}]{Camino_Goldman_PhysRevB_2005}%
  \BibitemOpen
  \bibfield  {author} {\bibinfo {author} {\bibfnamefont {F.~E.}\ \bibnamefont
  {Camino}}, \bibinfo {author} {\bibfnamefont {W.}~\bibnamefont {Zhou}}, \ and\
  \bibinfo {author} {\bibfnamefont {V.~J.}\ \bibnamefont {Goldman}},\ }\href
  {\doibase 10.1103/PhysRevB.72.075342} {\bibfield  {journal} {\bibinfo
  {journal} {Phys. Rev. B}\ }\textbf {\bibinfo {volume} {72}},\ \bibinfo
  {pages} {075342} (\bibinfo {year} {2005})}\BibitemShut {NoStop}%
\bibitem [{\citenamefont {Arovas}\ \emph {et~al.}(1984)\citenamefont {Arovas},
  \citenamefont {Schrieffer},\ and\ \citenamefont
  {Wilczek}}]{Arovas_Wilczek_PhysRevLett_1984}%
  \BibitemOpen
  \bibfield  {author} {\bibinfo {author} {\bibfnamefont {D.}~\bibnamefont
  {Arovas}}, \bibinfo {author} {\bibfnamefont {J.~R.}\ \bibnamefont
  {Schrieffer}}, \ and\ \bibinfo {author} {\bibfnamefont {F.}~\bibnamefont
  {Wilczek}},\ }\href {\doibase 10.1103/PhysRevLett.53.722} {\bibfield
  {journal} {\bibinfo  {journal} {Phys. Rev. Lett.}\ }\textbf {\bibinfo
  {volume} {53}},\ \bibinfo {pages} {722} (\bibinfo {year} {1984})}\BibitemShut
  {NoStop}%
\bibitem [{\citenamefont {Riccardo}\ \emph {et~al.}(2004)\citenamefont
  {Riccardo}, \citenamefont {Ramirez-Pastor},\ and\ \citenamefont
  {Rom\'a}}]{Riccardo_Roma_PhysRevLett_2004}%
  \BibitemOpen
  \bibfield  {author} {\bibinfo {author} {\bibfnamefont {J.~L.}\ \bibnamefont
  {Riccardo}}, \bibinfo {author} {\bibfnamefont {A.~J.}\ \bibnamefont
  {Ramirez-Pastor}}, \ and\ \bibinfo {author} {\bibfnamefont {F.}~\bibnamefont
  {Rom\'a}},\ }\href {\doibase 10.1103/PhysRevLett.93.186101} {\bibfield
  {journal} {\bibinfo  {journal} {Phys. Rev. Lett.}\ }\textbf {\bibinfo
  {volume} {93}},\ \bibinfo {pages} {186101} (\bibinfo {year}
  {2004})}\BibitemShut {NoStop}%
\bibitem [{\citenamefont {Riccardo}\ \emph {et~al.}(2005)\citenamefont
  {Riccardo}, \citenamefont {Rom\'a},\ and\ \citenamefont
  {Ramirez-Pastor}}]{Riccardo_Pastor_AppSurfSci_2005}%
  \BibitemOpen
  \bibfield  {author} {\bibinfo {author} {\bibfnamefont {J.}~\bibnamefont
  {Riccardo}}, \bibinfo {author} {\bibfnamefont {F.}~\bibnamefont {Rom\'a}}, \
  and\ \bibinfo {author} {\bibfnamefont {A.}~\bibnamefont {Ramirez-Pastor}},\
  }\href {\doibase 10.1016/j.apsusc.2005.02.067} {\bibfield  {journal}
  {\bibinfo  {journal} {Applied Surface Science}\ }\textbf {\bibinfo {volume}
  {252}},\ \bibinfo {pages} {505} (\bibinfo {year} {2005})},\ \bibinfo {note}
  {fifth International Symposium, Effects on surface heterogeneity in
  adsorption and catalysis on solids}\BibitemShut {NoStop}%
\bibitem [{\citenamefont {Cerofolini}(2006)}]{Cerofolini_JPhysA_2006}%
  \BibitemOpen
  \bibfield  {author} {\bibinfo {author} {\bibfnamefont {G.~F.}\ \bibnamefont
  {Cerofolini}},\ }\href {\doibase 10.1088/0305-4470/39/13/001} {\bibfield
  {journal} {\bibinfo  {journal} {Journal of Physics A: Mathematical and
  General}\ }\textbf {\bibinfo {volume} {39}},\ \bibinfo {pages} {3195}
  (\bibinfo {year} {2006})}\BibitemShut {NoStop}%
\bibitem [{\citenamefont {D\'avila}\ \emph
  {et~al.}(2009{\natexlab{a}})\citenamefont {D\'avila}, \citenamefont
  {Riccardo},\ and\ \citenamefont
  {Ramirez-Pastor}}]{Davila_Pastor_SurfSci_2009}%
  \BibitemOpen
  \bibfield  {author} {\bibinfo {author} {\bibfnamefont {M.}~\bibnamefont
  {D\'avila}}, \bibinfo {author} {\bibfnamefont {J.}~\bibnamefont {Riccardo}},
  \ and\ \bibinfo {author} {\bibfnamefont {A.}~\bibnamefont {Ramirez-Pastor}},\
  }\href {\doibase 10.1016/j.susc.2008.12.032} {\bibfield  {journal} {\bibinfo
  {journal} {Surface Science}\ }\textbf {\bibinfo {volume} {603}},\ \bibinfo
  {pages} {683} (\bibinfo {year} {2009}{\natexlab{a}})}\BibitemShut {NoStop}%
\bibitem [{\citenamefont {D\'avila}\ \emph
  {et~al.}(2009{\natexlab{b}})\citenamefont {D\'avila}, \citenamefont
  {Riccardo},\ and\ \citenamefont
  {Ramirez-Pastor}}]{Davila_Pastor_JChemPhys_2009}%
  \BibitemOpen
  \bibfield  {author} {\bibinfo {author} {\bibfnamefont {M.}~\bibnamefont
  {D\'avila}}, \bibinfo {author} {\bibfnamefont {J.~L.}\ \bibnamefont
  {Riccardo}}, \ and\ \bibinfo {author} {\bibfnamefont {A.~J.}\ \bibnamefont
  {Ramirez-Pastor}},\ }\href {\doibase 10.1063/1.3124163} {\bibfield  {journal}
  {\bibinfo  {journal} {The Journal of Chemical Physics}\ }\textbf {\bibinfo
  {volume} {130}},\ \bibinfo {pages} {174715} (\bibinfo {year}
  {2009}{\natexlab{b}})}\BibitemShut {NoStop}%
\bibitem [{\citenamefont {Matoz-Fernandez}\ \emph {et~al.}(2011)\citenamefont
  {Matoz-Fernandez}, \citenamefont {Linares},\ and\ \citenamefont
  {Ramirez-Pastor}}]{Fernandez_Pastor_Langmuir_2011}%
  \BibitemOpen
  \bibfield  {author} {\bibinfo {author} {\bibfnamefont {D.~A.}\ \bibnamefont
  {Matoz-Fernandez}}, \bibinfo {author} {\bibfnamefont {D.~H.}\ \bibnamefont
  {Linares}}, \ and\ \bibinfo {author} {\bibfnamefont {A.~J.}\ \bibnamefont
  {Ramirez-Pastor}},\ }\href {\doibase 10.1021/la104122h} {\bibfield  {journal}
  {\bibinfo  {journal} {Langmuir}\ }\textbf {\bibinfo {volume} {27}},\ \bibinfo
  {pages} {2456} (\bibinfo {year} {2011})}\BibitemShut {NoStop}%
\bibitem [{\citenamefont {Matoz-Fernandez}\ and\ \citenamefont
  {Ramirez-Pastor}(2014)}]{Fernandez_Pastor_ChemPhysLett_2014}%
  \BibitemOpen
  \bibfield  {author} {\bibinfo {author} {\bibfnamefont {D.}~\bibnamefont
  {Matoz-Fernandez}}\ and\ \bibinfo {author} {\bibfnamefont {A.}~\bibnamefont
  {Ramirez-Pastor}},\ }\href {\doibase 10.1016/j.cplett.2014.06.055} {\bibfield
   {journal} {\bibinfo  {journal} {Chemical Physics Letters}\ }\textbf
  {\bibinfo {volume} {610-611}},\ \bibinfo {pages} {131} (\bibinfo {year}
  {2014})}\BibitemShut {NoStop}%
\bibitem [{\citenamefont {Riccardo}\ \emph {et~al.}(2019)\citenamefont
  {Riccardo}, \citenamefont {Riccardo}, \citenamefont {Ramirez-Pastor},\ and\
  \citenamefont {Pasinetti}}]{Riccardo_Pasinetti_PhysRevLett_2019}%
  \BibitemOpen
  \bibfield  {author} {\bibinfo {author} {\bibfnamefont {J.~J.}\ \bibnamefont
  {Riccardo}}, \bibinfo {author} {\bibfnamefont {J.~L.}\ \bibnamefont
  {Riccardo}}, \bibinfo {author} {\bibfnamefont {A.~J.}\ \bibnamefont
  {Ramirez-Pastor}}, \ and\ \bibinfo {author} {\bibfnamefont {P.~M.}\
  \bibnamefont {Pasinetti}},\ }\href {\doibase 10.1103/PhysRevLett.123.020602}
  {\bibfield  {journal} {\bibinfo  {journal} {Phys. Rev. Lett.}\ }\textbf
  {\bibinfo {volume} {123}},\ \bibinfo {pages} {020602} (\bibinfo {year}
  {2019})}\BibitemShut {NoStop}%
\bibitem [{\citenamefont {Mezzasalma}(2019)}]{Mezzasalma_RSCAdv_2019}%
  \BibitemOpen
  \bibfield  {author} {\bibinfo {author} {\bibfnamefont {S.~A.}\ \bibnamefont
  {Mezzasalma}},\ }\href {\doibase 10.1039/C9RA02150G} {\bibfield  {journal}
  {\bibinfo  {journal} {RSC Adv.}\ }\textbf {\bibinfo {volume} {9}},\ \bibinfo
  {pages} {18678} (\bibinfo {year} {2019})}\BibitemShut {NoStop}%
\bibitem [{\citenamefont {Ihm}(1996)}]{Ihm_JPhysA_1996}%
  \BibitemOpen
  \bibfield  {author} {\bibinfo {author} {\bibfnamefont {J.}~\bibnamefont
  {Ihm}},\ }\href {\doibase 10.1088/0305-4470/29/1/001} {\bibfield  {journal}
  {\bibinfo  {journal} {Journal of Physics A: Mathematical and General}\
  }\textbf {\bibinfo {volume} {29}},\ \bibinfo {pages} {L1} (\bibinfo {year}
  {1996})}\BibitemShut {NoStop}%
\bibitem [{\citenamefont {Sharma}\ \emph {et~al.}(2015)\citenamefont {Sharma},
  \citenamefont {Reshetnyak}, \citenamefont {Andreev}, \citenamefont
  {Karbach},\ and\ \citenamefont {M\"uller}}]{Sharma_Muller_JStatMech_2015}%
  \BibitemOpen
  \bibfield  {author} {\bibinfo {author} {\bibfnamefont {G.~P.}\ \bibnamefont
  {Sharma}}, \bibinfo {author} {\bibfnamefont {Y.~K.}\ \bibnamefont
  {Reshetnyak}}, \bibinfo {author} {\bibfnamefont {O.~A.}\ \bibnamefont
  {Andreev}}, \bibinfo {author} {\bibfnamefont {M.}~\bibnamefont {Karbach}}, \
  and\ \bibinfo {author} {\bibfnamefont {G.}~\bibnamefont {M\"uller}},\ }\href
  {\doibase 10.1088/1742-5468/2015/01/p01034} {\bibfield  {journal} {\bibinfo
  {journal} {Journal of Statistical Mechanics: Theory and Experiment}\ }\textbf
  {\bibinfo {volume} {2015}},\ \bibinfo {pages} {P01034} (\bibinfo {year}
  {2015})}\BibitemShut {NoStop}%
\bibitem [{\citenamefont
  {Polychronakos}(1996{\natexlab{a}})}]{Polychronakos_PhysLettB_1996}%
  \BibitemOpen
  \bibfield  {author} {\bibinfo {author} {\bibfnamefont {A.~P.}\ \bibnamefont
  {Polychronakos}},\ }\href {\doibase 10.1016/0370-2693(95)01302-4} {\bibfield
  {journal} {\bibinfo  {journal} {Physics Letters B}\ }\textbf {\bibinfo
  {volume} {365}},\ \bibinfo {pages} {202} (\bibinfo {year}
  {1996}{\natexlab{a}})}\BibitemShut {NoStop}%
\bibitem [{\citenamefont
  {Polychronakos}(1996{\natexlab{b}})}]{Polychronakos_NuclPhysB_1996}%
  \BibitemOpen
  \bibfield  {author} {\bibinfo {author} {\bibfnamefont {A.~P.}\ \bibnamefont
  {Polychronakos}},\ }\href {\doibase 10.1016/0920-5632(95)00617-6} {\bibfield
  {journal} {\bibinfo  {journal} {Nuclear Physics B - Proceedings Supplements}\
  }\textbf {\bibinfo {volume} {45}},\ \bibinfo {pages} {81} (\bibinfo {year}
  {1996}{\natexlab{b}})},\ \bibinfo {note} {proceedings of the Trieste
  Conference on Recent Developments in Statistical Mechanics and Quantum Field
  Theory}\BibitemShut {NoStop}%
\bibitem [{\citenamefont {Chaturvedi}\ and\ \citenamefont
  {Srinivasan}(1997{\natexlab{a}})}]{Chaturvedi_Srinivasan_PhysicaA_1997}%
  \BibitemOpen
  \bibfield  {author} {\bibinfo {author} {\bibfnamefont {S.}~\bibnamefont
  {Chaturvedi}}\ and\ \bibinfo {author} {\bibfnamefont {V.}~\bibnamefont
  {Srinivasan}},\ }\href {\doibase 10.1016/S0378-4371(97)00348-8} {\bibfield
  {journal} {\bibinfo  {journal} {Physica A: Statistical Mechanics and its
  Applications}\ }\textbf {\bibinfo {volume} {246}},\ \bibinfo {pages} {576}
  (\bibinfo {year} {1997}{\natexlab{a}})}\BibitemShut {NoStop}%
\bibitem [{\citenamefont {Chaturvedi}\ and\ \citenamefont
  {Srinivasan}(1997{\natexlab{b}})}]{Chaturvedi_Srinivasan_PhysRevLett_1997}%
  \BibitemOpen
  \bibfield  {author} {\bibinfo {author} {\bibfnamefont {S.}~\bibnamefont
  {Chaturvedi}}\ and\ \bibinfo {author} {\bibfnamefont {V.}~\bibnamefont
  {Srinivasan}},\ }\href {\doibase 10.1103/PhysRevLett.78.4316} {\bibfield
  {journal} {\bibinfo  {journal} {Phys. Rev. Lett.}\ }\textbf {\bibinfo
  {volume} {78}},\ \bibinfo {pages} {4316} (\bibinfo {year}
  {1997}{\natexlab{b}})}\BibitemShut {NoStop}%
\bibitem [{\citenamefont {Murthy}\ and\ \citenamefont
  {Shankar}(1999)}]{Murthy_Shankar_PhysRevB_1999}%
  \BibitemOpen
  \bibfield  {author} {\bibinfo {author} {\bibfnamefont {M.~V.~N.}\
  \bibnamefont {Murthy}}\ and\ \bibinfo {author} {\bibfnamefont
  {R.}~\bibnamefont {Shankar}},\ }\href {\doibase 10.1103/PhysRevB.60.6517}
  {\bibfield  {journal} {\bibinfo  {journal} {Phys. Rev. B}\ }\textbf {\bibinfo
  {volume} {60}},\ \bibinfo {pages} {6517} (\bibinfo {year}
  {1999})}\BibitemShut {NoStop}%
\bibitem [{\citenamefont {Chung}\ and\ \citenamefont
  {Hassanabadi}(2018)}]{Chung_Hassanabadi_ModPhysLettB_2018}%
  \BibitemOpen
  \bibfield  {author} {\bibinfo {author} {\bibfnamefont {W.~S.}\ \bibnamefont
  {Chung}}\ and\ \bibinfo {author} {\bibfnamefont {H.}~\bibnamefont
  {Hassanabadi}},\ }\href {\doibase 10.1142/S0217984918500525} {\bibfield
  {journal} {\bibinfo  {journal} {Modern Physics Letters B}\ }\textbf {\bibinfo
  {volume} {32}},\ \bibinfo {pages} {1850052} (\bibinfo {year}
  {2018})}\BibitemShut {NoStop}%
\bibitem [{\citenamefont {Hoyuelos}(2018)}]{Hoyuelos_PhysicaA_2018}%
  \BibitemOpen
  \bibfield  {author} {\bibinfo {author} {\bibfnamefont {M.}~\bibnamefont
  {Hoyuelos}},\ }\href {\doibase 10.1016/j.physa.2017.09.006} {\bibfield
  {journal} {\bibinfo  {journal} {Physica A: Statistical Mechanics and its
  Applications}\ }\textbf {\bibinfo {volume} {490}},\ \bibinfo {pages} {944}
  (\bibinfo {year} {2018})}\BibitemShut {NoStop}%
\bibitem [{\citenamefont {Niven}(2009)}]{Niven_EurPhysJB_2009}%
  \BibitemOpen
  \bibfield  {author} {\bibinfo {author} {\bibfnamefont {R.~K.}\ \bibnamefont
  {Niven}},\ }\href {\doibase 10.1140/epjb/e2009-00168-5} {\bibfield  {journal}
  {\bibinfo  {journal} {The European Physical Journal B}\ }\textbf {\bibinfo
  {volume} {70}},\ \bibinfo {pages} {49} (\bibinfo {year} {2009})}\BibitemShut
  {NoStop}%
\bibitem [{\citenamefont {Niven}\ and\ \citenamefont
  {Grendar}(2009)}]{Niven_Grendar_PhysLettA_2009}%
  \BibitemOpen
  \bibfield  {author} {\bibinfo {author} {\bibfnamefont {R.~K.}\ \bibnamefont
  {Niven}}\ and\ \bibinfo {author} {\bibfnamefont {M.}~\bibnamefont
  {Grendar}},\ }\href {\doibase 10.1016/j.physleta.2008.12.025} {\bibfield
  {journal} {\bibinfo  {journal} {Physics Letters A}\ }\textbf {\bibinfo
  {volume} {373}},\ \bibinfo {pages} {621} (\bibinfo {year}
  {2009})}\BibitemShut {NoStop}%
\bibitem [{\citenamefont {Ourabah}\ and\ \citenamefont
  {Tribeche}(2018)}]{Ourabah_Tribeche_PhysRevE_2018}%
  \BibitemOpen
  \bibfield  {author} {\bibinfo {author} {\bibfnamefont {K.}~\bibnamefont
  {Ourabah}}\ and\ \bibinfo {author} {\bibfnamefont {M.}~\bibnamefont
  {Tribeche}},\ }\href {\doibase 10.1103/PhysRevE.97.032126} {\bibfield
  {journal} {\bibinfo  {journal} {Phys. Rev. E}\ }\textbf {\bibinfo {volume}
  {97}},\ \bibinfo {pages} {032126} (\bibinfo {year} {2018})}\BibitemShut
  {NoStop}%
\bibitem [{\citenamefont {Abutaleb}(2014)}]{Abutaleb_IntJTheoPhys_2014}%
  \BibitemOpen
  \bibfield  {author} {\bibinfo {author} {\bibfnamefont {A.~A.}\ \bibnamefont
  {Abutaleb}},\ }\href {\doibase 10.1007/s10773-014-2140-7} {\bibfield
  {journal} {\bibinfo  {journal} {International Journal of Theoretical
  Physics}\ }\textbf {\bibinfo {volume} {53}},\ \bibinfo {pages} {3893}
  (\bibinfo {year} {2014})}\BibitemShut {NoStop}%
\bibitem [{\citenamefont {Yan}(2021)}]{Yan_PhysRevE_2021}%
  \BibitemOpen
  \bibfield  {author} {\bibinfo {author} {\bibfnamefont {C.-T.}\ \bibnamefont
  {Yan}},\ }\href {\doibase 10.1103/PhysRevE.104.064118} {\bibfield  {journal}
  {\bibinfo  {journal} {Phys. Rev. E}\ }\textbf {\bibinfo {volume} {104}},\
  \bibinfo {pages} {064118} (\bibinfo {year} {2021})}\BibitemShut {NoStop}%
\bibitem [{\citenamefont {Kaniadakis}\ and\ \citenamefont
  {Quarati}(1993)}]{Kaniadakis_Quarati_PhysRevE_1993}%
  \BibitemOpen
  \bibfield  {author} {\bibinfo {author} {\bibfnamefont {G.}~\bibnamefont
  {Kaniadakis}}\ and\ \bibinfo {author} {\bibfnamefont {P.}~\bibnamefont
  {Quarati}},\ }\href {\doibase 10.1103/PhysRevE.48.4263} {\bibfield  {journal}
  {\bibinfo  {journal} {Phys. Rev. E}\ }\textbf {\bibinfo {volume} {48}},\
  \bibinfo {pages} {4263} (\bibinfo {year} {1993})}\BibitemShut {NoStop}%
\bibitem [{\citenamefont {Kaniadakis}\ and\ \citenamefont
  {Quarati}(1994)}]{Kaniadakis_Quarati_PhysRevE_1994}%
  \BibitemOpen
  \bibfield  {author} {\bibinfo {author} {\bibfnamefont {G.}~\bibnamefont
  {Kaniadakis}}\ and\ \bibinfo {author} {\bibfnamefont {P.}~\bibnamefont
  {Quarati}},\ }\href {\doibase 10.1103/PhysRevE.49.5103} {\bibfield  {journal}
  {\bibinfo  {journal} {Phys. Rev. E}\ }\textbf {\bibinfo {volume} {49}},\
  \bibinfo {pages} {5103} (\bibinfo {year} {1994})}\BibitemShut {NoStop}%
\bibitem [{\citenamefont {Kaniadakis}(1995)}]{Kaniadakis_PhysLettA_1995}%
  \BibitemOpen
  \bibfield  {author} {\bibinfo {author} {\bibfnamefont {G.}~\bibnamefont
  {Kaniadakis}},\ }\href {\doibase 10.1016/0375-9601(95)00414-X} {\bibfield
  {journal} {\bibinfo  {journal} {Physics Letters A}\ }\textbf {\bibinfo
  {volume} {203}},\ \bibinfo {pages} {229} (\bibinfo {year}
  {1995})}\BibitemShut {NoStop}%
\bibitem [{\citenamefont {Kaniadakis}\ \emph {et~al.}(1996)\citenamefont
  {Kaniadakis}, \citenamefont {Lavagno},\ and\ \citenamefont
  {Quarati}}]{Kaniadakis_Quarati_NuclPhysB_1996}%
  \BibitemOpen
  \bibfield  {author} {\bibinfo {author} {\bibfnamefont {G.}~\bibnamefont
  {Kaniadakis}}, \bibinfo {author} {\bibfnamefont {A.}~\bibnamefont {Lavagno}},
  \ and\ \bibinfo {author} {\bibfnamefont {P.}~\bibnamefont {Quarati}},\ }\href
  {\doibase 10.1016/0550-3213(96)00040-5} {\bibfield  {journal} {\bibinfo
  {journal} {Nuclear Physics B}\ }\textbf {\bibinfo {volume} {466}},\ \bibinfo
  {pages} {527} (\bibinfo {year} {1996})}\BibitemShut {NoStop}%
\bibitem [{\citenamefont {Kaniadakis}(2001)}]{Kaniadakis_PhysicaA_2001}%
  \BibitemOpen
  \bibfield  {author} {\bibinfo {author} {\bibfnamefont {G.}~\bibnamefont
  {Kaniadakis}},\ }\href {\doibase 10.1016/S0378-4371(01)00184-4} {\bibfield
  {journal} {\bibinfo  {journal} {Physica A: Statistical Mechanics and its
  Applications}\ }\textbf {\bibinfo {volume} {296}},\ \bibinfo {pages} {405}
  (\bibinfo {year} {2001})}\BibitemShut {NoStop}%
\bibitem [{\citenamefont {Kaniadakis}\ \emph {et~al.}(2002)\citenamefont
  {Kaniadakis}, \citenamefont {Quarati},\ and\ \citenamefont
  {Scarfone}}]{Kaniadakis_Scarfone_PhysicaA_2002}%
  \BibitemOpen
  \bibfield  {author} {\bibinfo {author} {\bibfnamefont {G.}~\bibnamefont
  {Kaniadakis}}, \bibinfo {author} {\bibfnamefont {P.}~\bibnamefont {Quarati}},
  \ and\ \bibinfo {author} {\bibfnamefont {A.}~\bibnamefont {Scarfone}},\
  }\href {\doibase 10.1016/S0378-4371(01)00643-4} {\bibfield  {journal}
  {\bibinfo  {journal} {Physica A: Statistical Mechanics and its Applications}\
  }\textbf {\bibinfo {volume} {305}},\ \bibinfo {pages} {76} (\bibinfo {year}
  {2002})},\ \bibinfo {note} {non Extensive Thermodynamics and Physical
  applications}\BibitemShut {NoStop}%
\bibitem [{\citenamefont {Kaniadakis}\ and\ \citenamefont
  {Scarfone}(2019)}]{Kaniadakis_Scarfone_Entropy_2019}%
  \BibitemOpen
  \bibfield  {author} {\bibinfo {author} {\bibfnamefont {G.}~\bibnamefont
  {Kaniadakis}}\ and\ \bibinfo {author} {\bibfnamefont {A.~M.}\ \bibnamefont
  {Scarfone}},\ }\href {\doibase 10.3390/e21090841} {\bibfield  {journal}
  {\bibinfo  {journal} {Entropy}\ }\textbf {\bibinfo {volume} {21}} (\bibinfo
  {year} {2019}),\ 10.3390/e21090841}\BibitemShut {NoStop}%
\bibitem [{\citenamefont {Medvedev}(1997)}]{Medvedev_PhysRevLett_1997}%
  \BibitemOpen
  \bibfield  {author} {\bibinfo {author} {\bibfnamefont {M.~V.}\ \bibnamefont
  {Medvedev}},\ }\href {\doibase 10.1103/PhysRevLett.78.4147} {\bibfield
  {journal} {\bibinfo  {journal} {Phys. Rev. Lett.}\ }\textbf {\bibinfo
  {volume} {78}},\ \bibinfo {pages} {4147} (\bibinfo {year}
  {1997})}\BibitemShut {NoStop}%
\bibitem [{\citenamefont {Meljanac}\ \emph {et~al.}(1999)\citenamefont
  {Meljanac}, \citenamefont {Milekovic},\ and\ \citenamefont
  {Ristic}}]{Meljanac_Ristic_ModPhysLettA_1999}%
  \BibitemOpen
  \bibfield  {author} {\bibinfo {author} {\bibfnamefont {S.}~\bibnamefont
  {Meljanac}}, \bibinfo {author} {\bibfnamefont {M.}~\bibnamefont {Milekovic}},
  \ and\ \bibinfo {author} {\bibfnamefont {R.}~\bibnamefont {Ristic}},\ }\href
  {\doibase 10.1142/S0217732399002509} {\bibfield  {journal} {\bibinfo
  {journal} {Modern Physics Letters A}\ }\textbf {\bibinfo {volume} {14}},\
  \bibinfo {pages} {2413} (\bibinfo {year} {1999})}\BibitemShut {NoStop}%
\bibitem [{\citenamefont {Berndt}\ \emph {et~al.}(1983)\citenamefont {Berndt},
  \citenamefont {Evans},\ and\ \citenamefont
  {Wilson}}]{Berndt_Wilson_AdvMath_1983}%
  \BibitemOpen
  \bibfield  {author} {\bibinfo {author} {\bibfnamefont {B.~C.}\ \bibnamefont
  {Berndt}}, \bibinfo {author} {\bibfnamefont {R.~J.}\ \bibnamefont {Evans}}, \
  and\ \bibinfo {author} {\bibfnamefont {B.}~\bibnamefont {Wilson}},\ }\href
  {\doibase https://doi.org/10.1016/0001-8708(83)90071-3} {\bibfield  {journal}
  {\bibinfo  {journal} {Advances in Mathematics}\ }\textbf {\bibinfo {volume}
  {49}},\ \bibinfo {pages} {123} (\bibinfo {year} {1983})}\BibitemShut
  {NoStop}%
\bibitem [{sci(v611)}]{scilab}%
  \BibitemOpen
  \href {www.scilab.org} {\emph {\bibinfo {title} {Scilab}}}\ (\bibinfo {year}
  {v6.1.1})\ \bibinfo {note} {copyright © 1989-2005 INRIA ENPC.}\BibitemShut
  {Stop}%
\bibitem [{\citenamefont {Bhaduri}\ \emph {et~al.}(2010)\citenamefont
  {Bhaduri}, \citenamefont {Murthy},\ and\ \citenamefont
  {Sen}}]{Bhaduri_Sen_JPhysA_2010}%
  \BibitemOpen
  \bibfield  {author} {\bibinfo {author} {\bibfnamefont {R.~K.}\ \bibnamefont
  {Bhaduri}}, \bibinfo {author} {\bibfnamefont {M.~V.~N.}\ \bibnamefont
  {Murthy}}, \ and\ \bibinfo {author} {\bibfnamefont {D.}~\bibnamefont {Sen}},\
  }\href {\doibase 10.1088/1751-8113/43/4/045002} {\bibfield  {journal}
  {\bibinfo  {journal} {Journal of Physics A: Mathematical and Theoretical}\
  }\textbf {\bibinfo {volume} {43}},\ \bibinfo {pages} {045002} (\bibinfo
  {year} {2010})}\BibitemShut {NoStop}%
\bibitem [{\citenamefont {Hill}(1962)}]{Hill_JChemPhys_1962}%
  \BibitemOpen
  \bibfield  {author} {\bibinfo {author} {\bibfnamefont {T.~L.}\ \bibnamefont
  {Hill}},\ }\href {\doibase 10.1063/1.1732447} {\bibfield  {journal} {\bibinfo
   {journal} {The Journal of Chemical Physics}\ }\textbf {\bibinfo {volume}
  {36}},\ \bibinfo {pages} {3182} (\bibinfo {year} {1962})}\BibitemShut
  {NoStop}%
\bibitem [{\citenamefont {Hill}(2001{\natexlab{a}})}]{Hill_NanoLett_2001}%
  \BibitemOpen
  \bibfield  {author} {\bibinfo {author} {\bibfnamefont {T.~L.}\ \bibnamefont
  {Hill}},\ }\href {\doibase 10.1021/nl010010d} {\bibfield  {journal} {\bibinfo
   {journal} {Nano Letters}\ }\textbf {\bibinfo {volume} {1}},\ \bibinfo
  {pages} {111} (\bibinfo {year} {2001}{\natexlab{a}})}\BibitemShut {NoStop}%
\bibitem [{\citenamefont {Hill}(2001{\natexlab{b}})}]{Hill_NanoLett_2001_1}%
  \BibitemOpen
  \bibfield  {author} {\bibinfo {author} {\bibfnamefont {T.~L.}\ \bibnamefont
  {Hill}},\ }\href {\doibase 10.1021/nl010027w} {\bibfield  {journal} {\bibinfo
   {journal} {Nano Letters}\ }\textbf {\bibinfo {volume} {1}},\ \bibinfo
  {pages} {273} (\bibinfo {year} {2001}{\natexlab{b}})}\BibitemShut {NoStop}%
\bibitem [{\citenamefont {Hill}(2001{\natexlab{c}})}]{Hill_NanoLett_2001_2}%
  \BibitemOpen
  \bibfield  {author} {\bibinfo {author} {\bibfnamefont {T.~L.}\ \bibnamefont
  {Hill}},\ }\href {\doibase 10.1021/nl010009e} {\bibfield  {journal} {\bibinfo
   {journal} {Nano Letters}\ }\textbf {\bibinfo {volume} {1}},\ \bibinfo
  {pages} {159} (\bibinfo {year} {2001}{\natexlab{c}})}\BibitemShut {NoStop}%
\bibitem [{\citenamefont {Chamberlin}(2015)}]{Chamberlin_Entropy_2015}%
  \BibitemOpen
  \bibfield  {author} {\bibinfo {author} {\bibfnamefont {R.~V.}\ \bibnamefont
  {Chamberlin}},\ }\href {\doibase 10.3390/e17010052} {\bibfield  {journal}
  {\bibinfo  {journal} {Entropy}\ }\textbf {\bibinfo {volume} {17}},\ \bibinfo
  {pages} {52} (\bibinfo {year} {2015})}\BibitemShut {NoStop}%
\bibitem [{\citenamefont {Chamberlin}\ \emph {et~al.}(2021)\citenamefont
  {Chamberlin}, \citenamefont {Clark}, \citenamefont {Mujica},\ and\
  \citenamefont {Wolf}}]{Chamberlin_Wolf_Symmetry_2021}%
  \BibitemOpen
  \bibfield  {author} {\bibinfo {author} {\bibfnamefont {R.~V.}\ \bibnamefont
  {Chamberlin}}, \bibinfo {author} {\bibfnamefont {M.~R.}\ \bibnamefont
  {Clark}}, \bibinfo {author} {\bibfnamefont {V.}~\bibnamefont {Mujica}}, \
  and\ \bibinfo {author} {\bibfnamefont {G.~H.}\ \bibnamefont {Wolf}},\ }\href
  {\doibase 10.3390/sym13040721} {\bibfield  {journal} {\bibinfo  {journal}
  {Symmetry}\ }\textbf {\bibinfo {volume} {13}} (\bibinfo {year} {2021}),\
  10.3390/sym13040721}\BibitemShut {NoStop}%
\bibitem [{\citenamefont {Anghel}(2008)}]{Anghel_PhysLettA_2008}%
  \BibitemOpen
  \bibfield  {author} {\bibinfo {author} {\bibfnamefont {D.-V.}\ \bibnamefont
  {Anghel}},\ }\href {\doibase https://doi.org/10.1016/j.physleta.2008.07.044}
  {\bibfield  {journal} {\bibinfo  {journal} {Physics Letters A}\ }\textbf
  {\bibinfo {volume} {372}},\ \bibinfo {pages} {5745} (\bibinfo {year}
  {2008})}\BibitemShut {NoStop}%
\bibitem [{\citenamefont {Anghel}(2012)}]{Anghel_PhysScr_2012}%
  \BibitemOpen
  \bibfield  {author} {\bibinfo {author} {\bibfnamefont {D.-V.}\ \bibnamefont
  {Anghel}},\ }\href {\doibase 10.1088/0031-8949/2012/t151/014079} {\bibfield
  {journal} {\bibinfo  {journal} {Physica Scripta}\ }\textbf {\bibinfo {volume}
  {T151}},\ \bibinfo {pages} {014079} (\bibinfo {year} {2012})}\BibitemShut
  {NoStop}%
\bibitem [{\citenamefont {Lichtenstein}\ \emph {et~al.}(2012)\citenamefont
  {Lichtenstein}, \citenamefont {B\"uchner}, \citenamefont {Yang},
  \citenamefont {Shaikhutdinov}, \citenamefont {Heyde}, \citenamefont {Sierka},
  \citenamefont {W\l{}odarczyk}, \citenamefont {Sauer},\ and\ \citenamefont
  {Freund}}]{Lichtenstein_Freund_AngewChem_2012}%
  \BibitemOpen
  \bibfield  {author} {\bibinfo {author} {\bibfnamefont {L.}~\bibnamefont
  {Lichtenstein}}, \bibinfo {author} {\bibfnamefont {C.}~\bibnamefont
  {B\"uchner}}, \bibinfo {author} {\bibfnamefont {B.}~\bibnamefont {Yang}},
  \bibinfo {author} {\bibfnamefont {S.}~\bibnamefont {Shaikhutdinov}}, \bibinfo
  {author} {\bibfnamefont {M.}~\bibnamefont {Heyde}}, \bibinfo {author}
  {\bibfnamefont {M.}~\bibnamefont {Sierka}}, \bibinfo {author} {\bibfnamefont
  {R.}~\bibnamefont {W\l{}odarczyk}}, \bibinfo {author} {\bibfnamefont
  {J.}~\bibnamefont {Sauer}}, \ and\ \bibinfo {author} {\bibfnamefont {H.-J.}\
  \bibnamefont {Freund}},\ }\href {\doibase 10.1002/anie.201107097} {\bibfield
  {journal} {\bibinfo  {journal} {Angewandte Chemie International Edition}\
  }\textbf {\bibinfo {volume} {51}},\ \bibinfo {pages} {404} (\bibinfo {year}
  {2012})}\BibitemShut {NoStop}%
\bibitem [{\citenamefont {Roy}\ \emph {et~al.}(2018)\citenamefont {Roy},
  \citenamefont {Heyde},\ and\ \citenamefont
  {Heuer}}]{Roy_Heuer_PhysChemChemPhys_2018}%
  \BibitemOpen
  \bibfield  {author} {\bibinfo {author} {\bibfnamefont {P.~K.}\ \bibnamefont
  {Roy}}, \bibinfo {author} {\bibfnamefont {M.}~\bibnamefont {Heyde}}, \ and\
  \bibinfo {author} {\bibfnamefont {A.}~\bibnamefont {Heuer}},\ }\href
  {\doibase 10.1039/C8CP01313F} {\bibfield  {journal} {\bibinfo  {journal}
  {Phys. Chem. Chem. Phys.}\ }\textbf {\bibinfo {volume} {20}},\ \bibinfo
  {pages} {14725} (\bibinfo {year} {2018})}\BibitemShut {NoStop}%
\bibitem [{\citenamefont {Roy}\ and\ \citenamefont
  {Heuer}(2019{\natexlab{a}})}]{Roy_Heuer_PhysRevLett_2019}%
  \BibitemOpen
  \bibfield  {author} {\bibinfo {author} {\bibfnamefont {P.~K.}\ \bibnamefont
  {Roy}}\ and\ \bibinfo {author} {\bibfnamefont {A.}~\bibnamefont {Heuer}},\
  }\href {\doibase 10.1103/PhysRevLett.122.016104} {\bibfield  {journal}
  {\bibinfo  {journal} {Phys. Rev. Lett.}\ }\textbf {\bibinfo {volume} {122}},\
  \bibinfo {pages} {016104} (\bibinfo {year} {2019}{\natexlab{a}})}\BibitemShut
  {NoStop}%
\bibitem [{\citenamefont {Roy}\ and\ \citenamefont
  {Heuer}(2019{\natexlab{b}})}]{Roy_Heuer_JPhysCondMat_2019}%
  \BibitemOpen
  \bibfield  {author} {\bibinfo {author} {\bibfnamefont {P.~K.}\ \bibnamefont
  {Roy}}\ and\ \bibinfo {author} {\bibfnamefont {A.}~\bibnamefont {Heuer}},\
  }\href {\doibase 10.1088/1361-648x/ab0a13} {\bibfield  {journal} {\bibinfo
  {journal} {Journal of Physics: Condensed Matter}\ }\textbf {\bibinfo {volume}
  {31}},\ \bibinfo {pages} {225703} (\bibinfo {year}
  {2019}{\natexlab{b}})}\BibitemShut {NoStop}%
\end{thebibliography}
\end{document}